\documentclass{aa}

\usepackage{graphicx}
\usepackage{natbib}
\usepackage{color}
\usepackage{float}
\usepackage{amsmath}
\usepackage{txfonts}
\usepackage{pdflscape}

\newcommand{\mlc}{\multicolumn{1}{c}}

\begin{document}

\title{First apsidal motion and light curve analysis \\ of 162 eccentric eclipsing binaries from LMC}

\author{Zasche, P.~\inst{1},
     Wolf, M.~\inst{1},
     Ku\v{c}\'akov\'a, H.~\inst{1,2,3},
     K\'ara, J.~\inst{1},
     Merc, J.~\inst{1,4},
     Zejda, M.~\inst{5},
     Skarka, M.~\inst{5, 2},
     Jan\'{\i}k, J.~\inst{5},
     Kurf\"urst, P.~\inst{5}.
}

\offprints{Petr Zasche, \email{zasche@sirrah.troja.mff.cuni.cz}}

 \institute{
  $^{1}$ Astronomical Institute, Charles University, Faculty of Mathematics and Physics, CZ-180~00, Praha 8, V~Hole\v{s}ovi\v{c}k\'ach 2, Czech Republic\\
  $^{2}$ Astronomical Institute, Academy of Sciences, Fri\v{c}ova 298, CZ-251 65, Ond\v{r}ejov, Czech Republic\\
  $^{3}$ Research Centre for Theoretical Physics and Astrophysics, Institute of Physics, Silesian University in Opava, Bezru\v{c}ovo n\'am. 13, CZ-746 01, Opava, Czech Republic\\
  $^{4}$ Institute of Physics, Faculty of Science, P. J. \v{S}af{\'a}rik University, Park Angelinum 9, 040 01 Ko\v{s}ice, Slovakia \\
  $^{5}$ Department of Theoretical Physics and Astrophysics, Masaryk University, Kotl\'a\v{r}sk\'a 2, CZ-611~37, Brno, Czech Republic }

\titlerunning{Study of 162 LMC eccentric eclipsing binaries}
\authorrunning{Zasche et al.}

  \date{Received \today; accepted ???}

\abstract{We present an extensive study of 162 early-type binary systems located in the LMC galaxy
that show apsidal motion and have never been studied before. For the sample systems, we performed
light curve and apsidal motion modelling for the first time. These systems have a median orbital
period of 2.2\,days and typical periods of the apsidal motion were derived to be of the order of
decades. We identified two record-breaking systems. The first, OGLE LMC-ECL-22613, shows the
shortest known apsidal motion period among systems with main sequence components (6.6\,years); it
contains a third component with an orbital period of 23\,years. The second, OGLE LMC-ECL-17226, is
an eccentric system with the shortest known orbital period (0.9879\,days) and with quite fast
apsidal motion period (11\,years). Among the studied systems, 36 new triple-star candidates were
identified based on the additional period variations. This represents more than 20\,\% of all
studied systems, which is in agreement with the statistics of multiples in our Galaxy. However, the
fraction should only be considered as a lower limit of these early-type stars in the LMC because of
our method of detection, data coverage, and limited precision of individual times of eclipses. }


\maketitle

\section{Introduction} \label{intro}

The use of eclipsing binaries (EBs) for the precise derivation of the physical parameters of their
components, such as masses and radii, still represents the most powerful method for deriving these
quantities (see e.g. \citealt{2012ocpd.conf...51S}). And thanks to these parameters we are able to
improve and calibrate the existing models of stellar structure and evolution
\citep{2010A&ARv..18...67T}. We can also use the EBs as independent distance indicators, even
outside of our Galaxy (e.g. \citealt{2005MNRAS.357..304H}, or \citealt{2010A&A...509A..70V}).

Moreover, the eccentric EBs with detectable apsidal motion are important objects for studies of the
internal structure constants and also to test the general relativity \citep{1993A&A...277..487C}.
For the systems, where the presence of an additional third body was detected,  the more interesting
dynamical interactions were also studied by \cite{2015MNRAS.448..946B}, among others.
 A large compilation of 623 galactic eccentric EBs was presented by
\cite{2018ApJS..235...41K}. The authors showed that about 5\% of the systems they analysed also
have a third unseen component (only based on eclipse timing variations, hereafter ETV). Finally,
the effects like orbital circularisation \citep{2008EAS....29...67Z} can also be studied by means
of distribution of systems in the period-eccentricity diagram (see e.g.
\citealt{2012ApJ...751....4K}, or \citealt{2005ApJ...620..970M}).

Several similar studies were published on this topic, focusing on  the eccentric EBs in the LMC and
SMC and yielding the apsidal motion parameters, but the present study is  the most extensive one.
The studies that  use  a similar method of analysing the photometry and yield  the apsidal motion,
and that contain three or more systems from the  LMC and/or SMC, are summarised in Table
\ref{LMCSMCEEBs}.

\begin{table}
\caption{Compilation of papers studying LMC/SMC eccentric binaries.}  \label{LMCSMCEEBs}
  \centering
\begin{tabular}{lcc}
  \hline\noalign{\smallskip}
\multicolumn{1}{c}{Paper} & Field & \multicolumn{1}{c}{Number of systems}\\\noalign{\smallskip}
  \hline\noalign{\smallskip}
 \cite{2013AaA...558A..51Z} & LMC   & 5  \\
 \cite{2014AaA...572A..71Z} & SMC   & 18 \\
 \cite{2015AJ....150..183Z} & LMC   & 13 \\
 \cite{2015AJ....150....1H} & SMC   & 27 \\
 \cite{2016MNRAS.460..650H} & SMC   & 90 \\
 \cite{2019AJ....158..185H} & LMC   &  3 \\
 \cite{2019AJ....157...87Z} & SMC   & 21 \\
 Our present study          & LMC   &162 \\
 \noalign{\smallskip}\hline
\end{tabular}
\end{table}

\section{Method of  analysis}  \label{methods}

The selection of potential apsidal motion systems was  a  byproduct of our analysis of the whole
LMC catalogue of EBs scanned for doubly eclipsing systems. We identified several hundreds of
prospective eccentric systems for a more detailed analysis, but only a small part of them were
found to fit our criteria. With these limitations we mainly mean the apsidal motion period, which
has to be shorter than 200 years (principal limit due to a typical data coverage, see below). This
was the strictest criterion because a large portion of the systems show apsidal motion but with
very long periods. Other criteria were the data coverage  of the light curve (LC)  and its
photometric amplitude (depth of minima), but also good coverage of the apsidal motion in the $O-C$
diagram, where the apsidal motion should be unambiguously detectable.

For our whole analysis we only used the photometric data. Databases like OGLE and MACHO were used
when available, while for some systems  new photometry from the Danish 1.54 m telescope (hereafter
DK154) was also obtained\footnote{\tiny{Also available online via CDS.}}. MACHO data
\citep{1997ApJ...486..697A} cover the period 1992-2000, while the OGLE experiment
\citep{1992AcA....42..253U}  was run in several parts: OGLE II in 1997-2000, OGLE III 2001-2009,
and OGLE IV 2010-2014 for both Magellanic Clouds. The Photometric Data Retriever PDR
\citep{2019CoSka..49..132Z} was used for some systems to obtain the data.

The LC analysis using the {\sc PHOEBE 0.32svn} code \citep{2005ApJ...628..426P} (which is based on
the Wilson-Devinney algorithm \citealt{1971ApJ...166..605W}) was routinely used for the analysis of
all systems in our sample. However, when there was no spectroscopy, several assumptions had to be
made: the mass ratio was fixed at $q=1.0$, and the coefficients of albedo $A$, limb-darkening $x$,
and gravity brightening $g$ were kept fixed at their suggested values for hot stars. The quality of
LCs does not usually allow us to also fit  these second-order parameters.

A necessary input parameter is the primary temperature, which was estimated from the de-reddened
photometric index $(B-V)_0$, which was derived following the procedure by
\cite{1958LowOB...4...37J} using the values of photometric indices $(B-V)$ and $(U-B)$. The use of
this method is substantiated because we deal here only with hot stars  (of spectral types O-B-A),
for which this presented $(B-V)_0$ transformation works. These values are given below in Tables
\ref{InfoSystems} and \ref{InfoSystems2}. The primary temperature was derived from the $(B-V)_0$
photometric index using the assumption of main-sequence components and the tables by
\cite{2013ApJS..208....9P} and its later updates available
online\footnote{\tiny{www.pas.rochester.edu/$\sim$emamajek/EEM$\_$dwarf$\_$UBVIJHK$\_$colors$\_$Teff.txt}}.

This procedure of $T_1$ estimation was used for majority of stars, but for 27 systems from our
sample this resulted in rather dubious results. From these 27 systems,   19 systems were found with
very low values of their indices $(B-V)_0 < -0.31$. Normal O-type stars should have the index in
the range -0.32 to -0.31, but here we deal with almost all systems having much lower unrealistic
values of this photometric index. Therefore, we decided to skip this colour  information
completely. Moreover,  four
  systems have unavailable indices from \cite{2002ApJS..141...81M} and \cite{2002AJ....123..855Z}; two
systems have unrealistically high values of $(B-V)_0 > 0$; and finally two systems have values of
their $(B-V)_0$ indices that are too different from those of  \cite{2002ApJS..141...81M} and
\cite{2002AJ....123..855Z}. For these 27 systems we only roughly estimated their effective
temperatures from their apparent magnitudes (and this magnitude-temperature relation was
constructed using the other systems with realistic $(B-V)_0$ values). These stars are flagged in
Tables \ref{InfoSystems} and \ref{InfoSystems2} with asterisks. According to this temperature
estimation it seems that all of the stars have  effective temperatures below 32000~K, hence none of
our stars is an O-type star.

From the complete LC analysis, we constructed an LC template, which was later used to derive the
times of eclipses via our AFP method (see \citealt{2014AaA...572A..71Z}). For the sake of brevity,
all the figures of the LCs given here were plotted  during a shorter time interval to avoid phase
smearing due to change of argument of periastron $\omega$   (and using mostly only OGLE III data in
the $I$ filter).

At this point it would be useful to explain our method step by step. First, we only roughly
estimated the orbital period and the eccentricity. With this assumption, we made the LC template
and performed a preliminary period analysis resulting in better estimation of the true eccentricity
and the period. With these values of eccentricity and period, we constructed a better LC fit. With
the new LC template we were able to perform better apsidal motion analysis. With this iterative
procedure we were able to get the final solution, which is the one presented in Tables
\ref{LCOCparam} and \ref{LCOCparam2}. It should also be mentioned that for the LC analysis we used
the photometric data phased only for some shorter time interval (typically one or two OGLE seasons
only, i.e. 1 or 2 years) to eliminate the effect of apsidal motion causing the change of duration
and depth of the LC for the template. For systems where the apsidal motion is fast, we constructed
several such templates and used them separately for certain time intervals of our photometric data.
The rate of apsidal motion was derived from the eclipse times, which we found is better than that
derived from the LC fits. The apsidal motion analysis uses the standard technique of least-squares
minimisation, as introduced   by \cite{1983Ap&SS..92..203G} and \cite{1995Ap&SS.226...99G}, among
others.

\begin{figure*}
  \centering
  \includegraphics[width=0.865\textwidth]{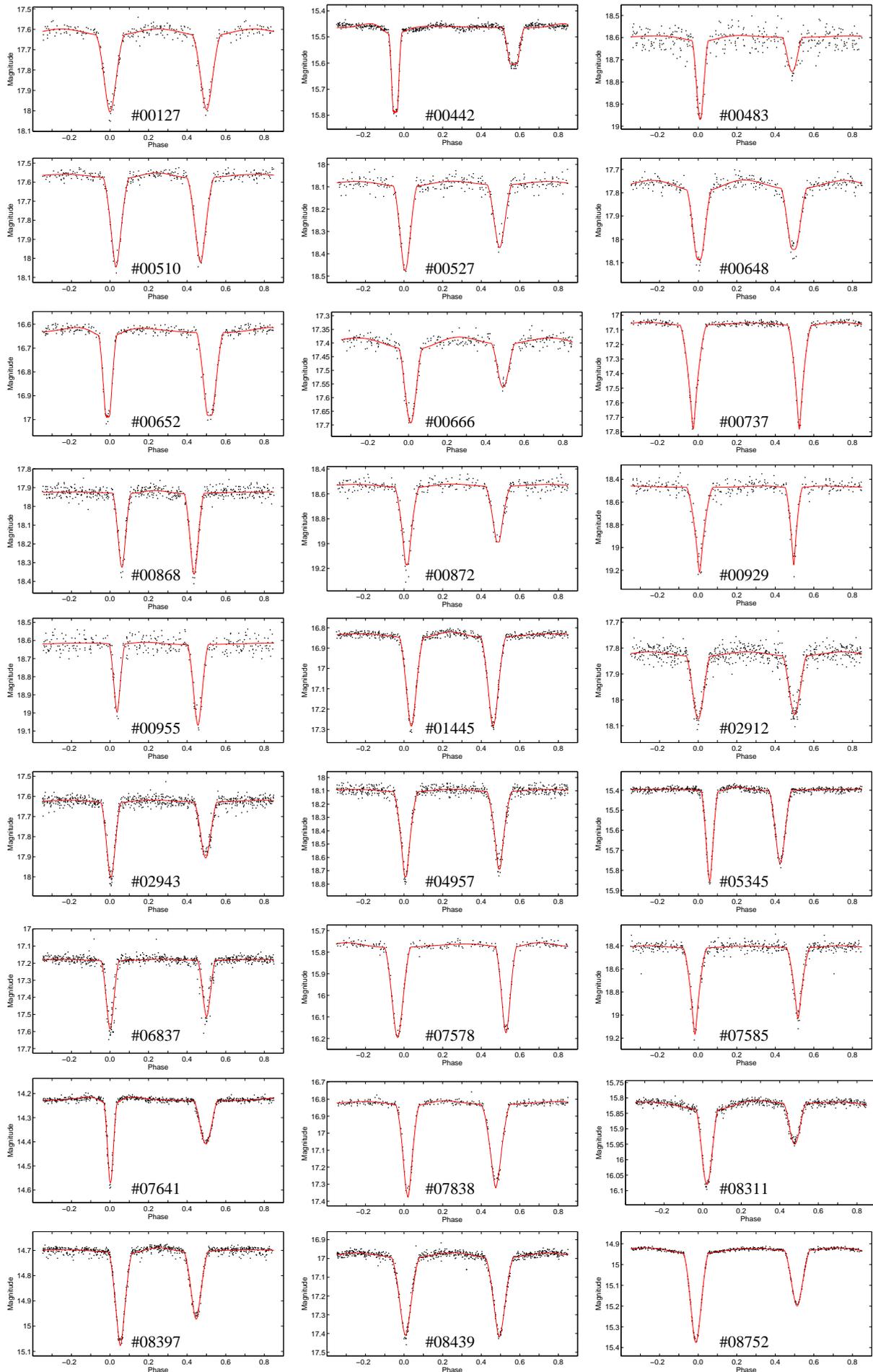}
  \caption{Plot of the light curves, part 1.}
  \label{FigLC1}
\end{figure*}

\begin{figure*}
  \centering
  \includegraphics[width=0.865\textwidth]{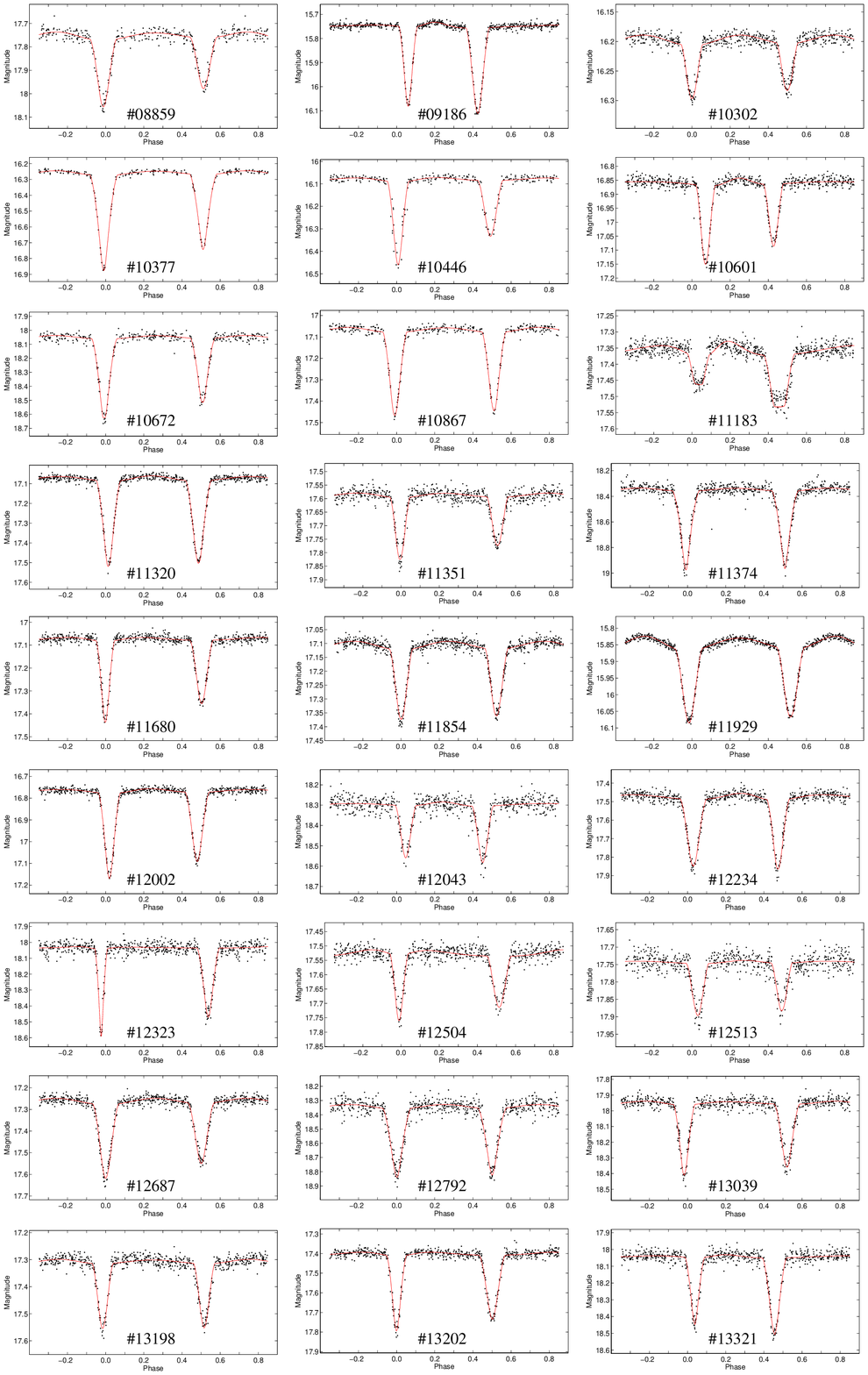}
  \caption{Plot of the light curves, part 2.}
  \label{FigLC2}
\end{figure*}

\begin{figure*}
  \centering
  \includegraphics[width=0.865\textwidth]{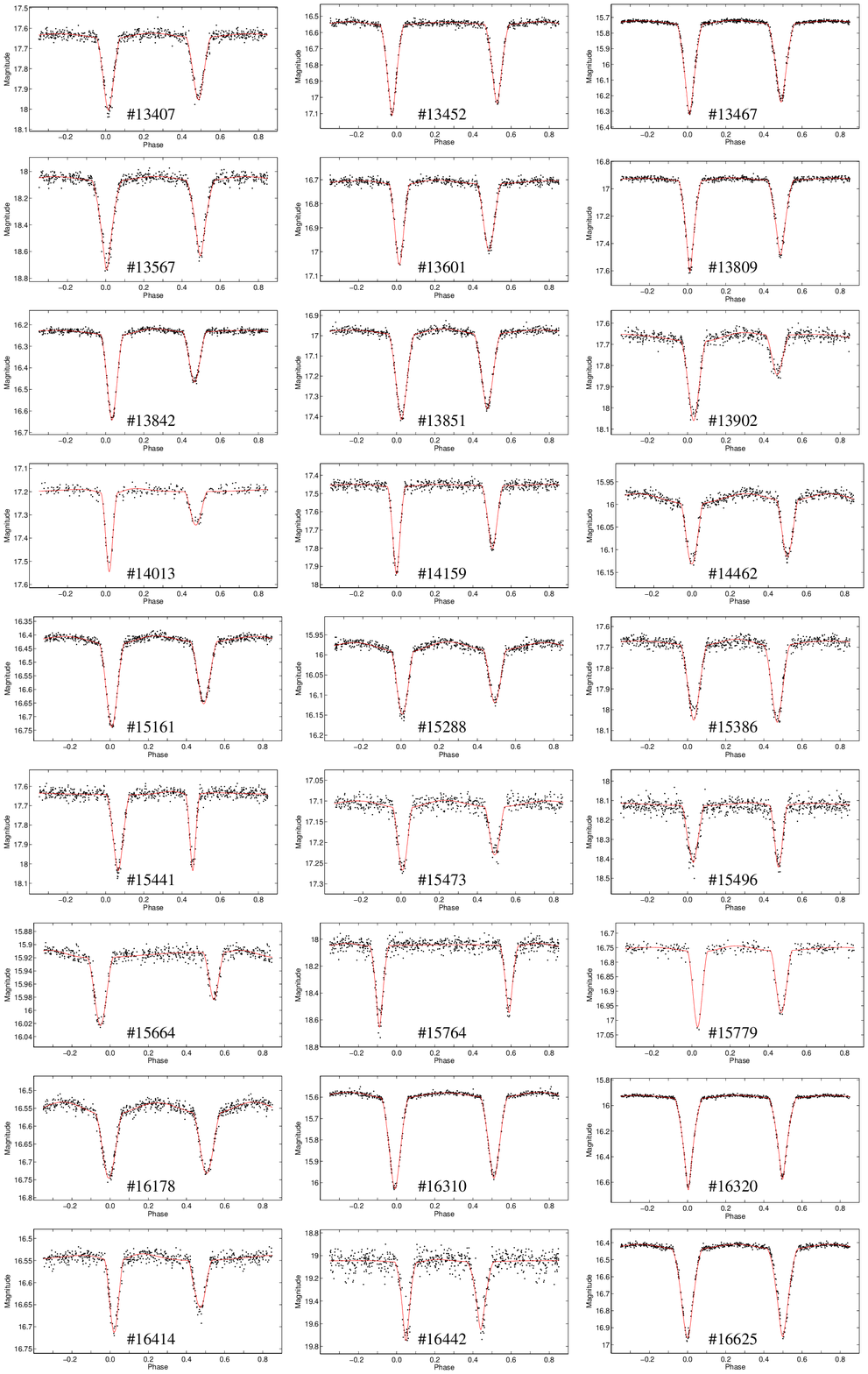}
  \caption{Plot of the light curves, part 3.}
  \label{FigLC3}
\end{figure*}

\begin{figure*}
  \centering
  \includegraphics[width=0.865\textwidth]{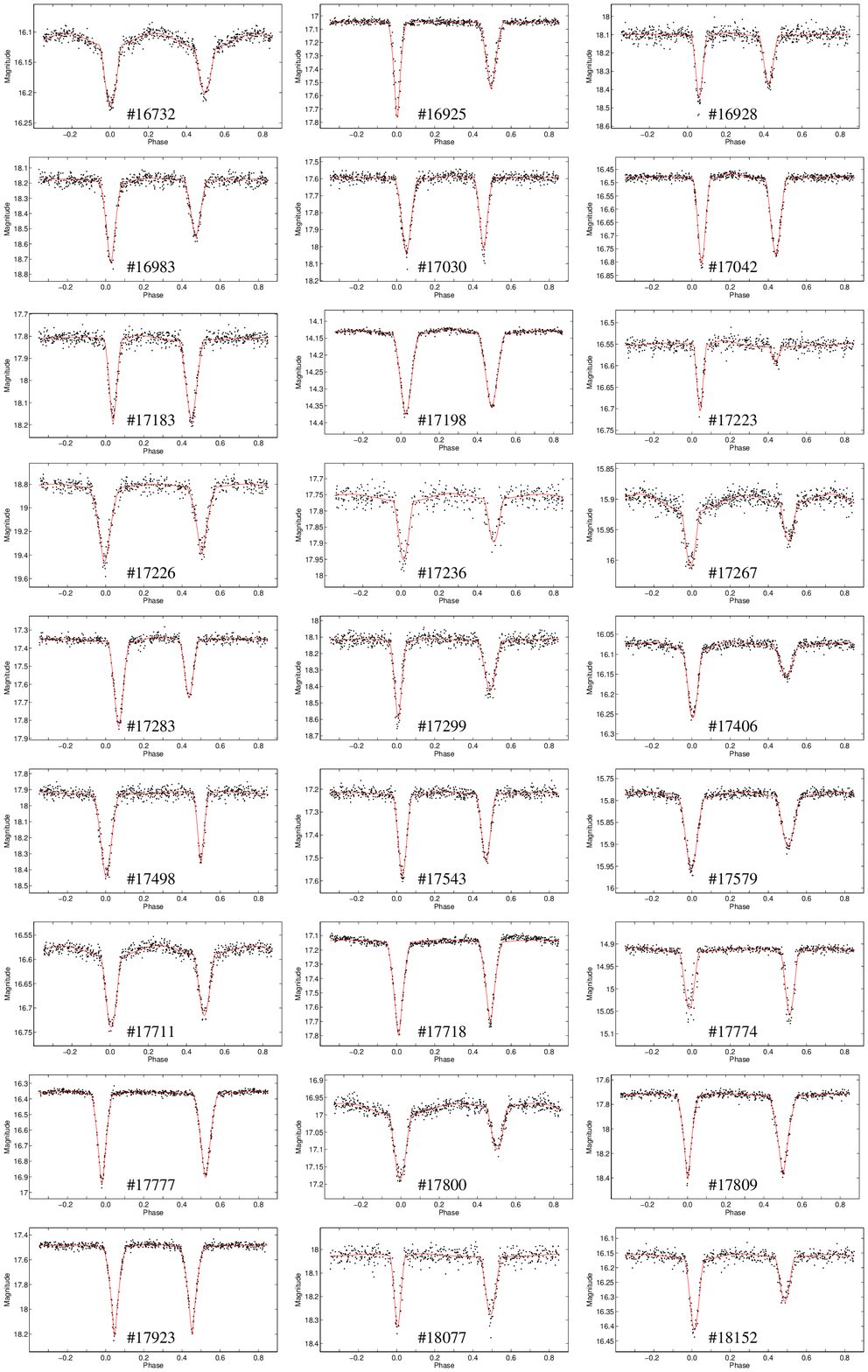}
  \caption{Plot of the light curves, part 4.}
  \label{FigLC4}
\end{figure*}

\begin{figure*}
  \centering
  \includegraphics[width=0.865\textwidth]{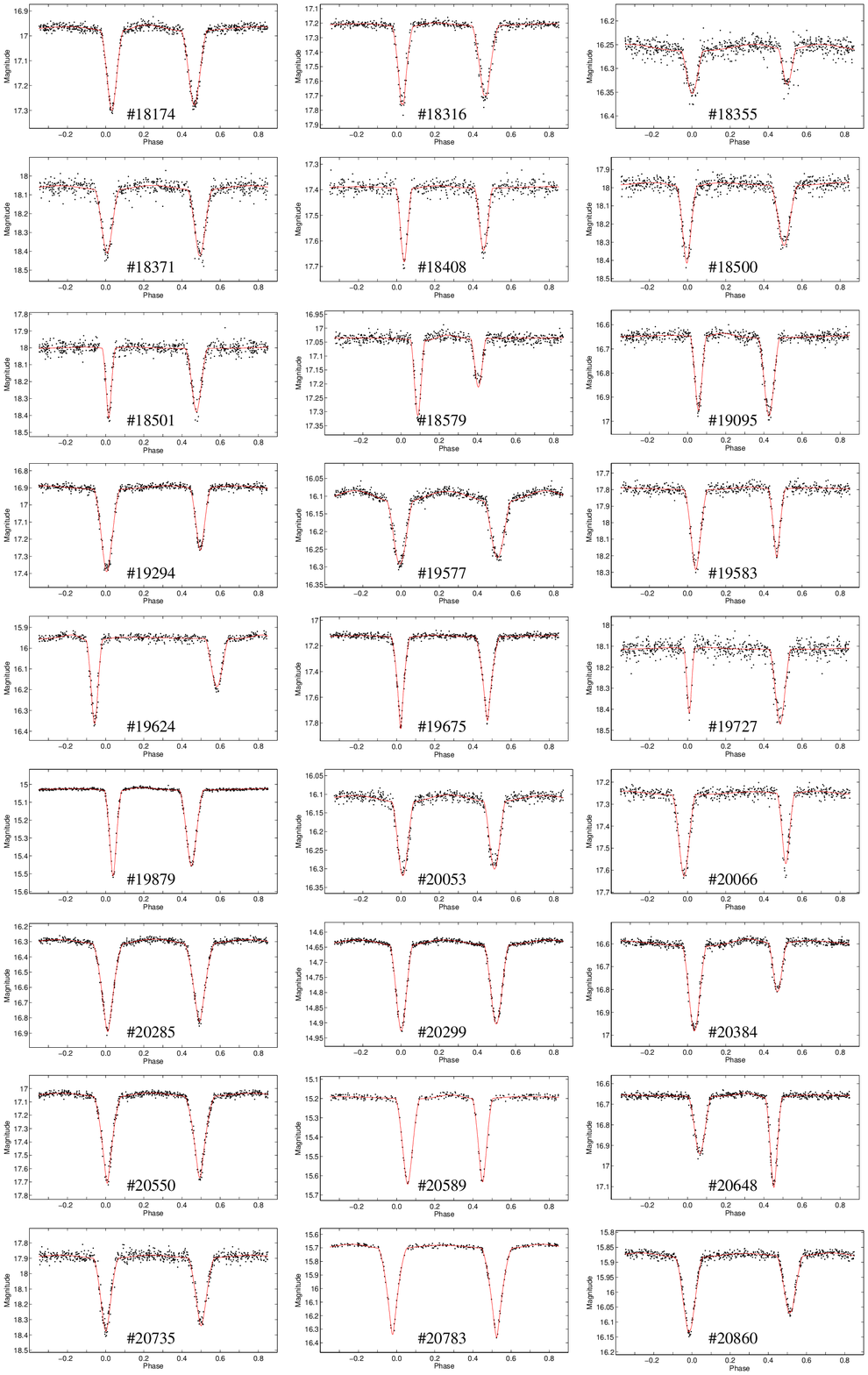}
  \caption{Plot of the light curves, part 5.}
  \label{FigLC5}
\end{figure*}

\begin{figure*}
  \centering
  \includegraphics[width=0.865\textwidth]{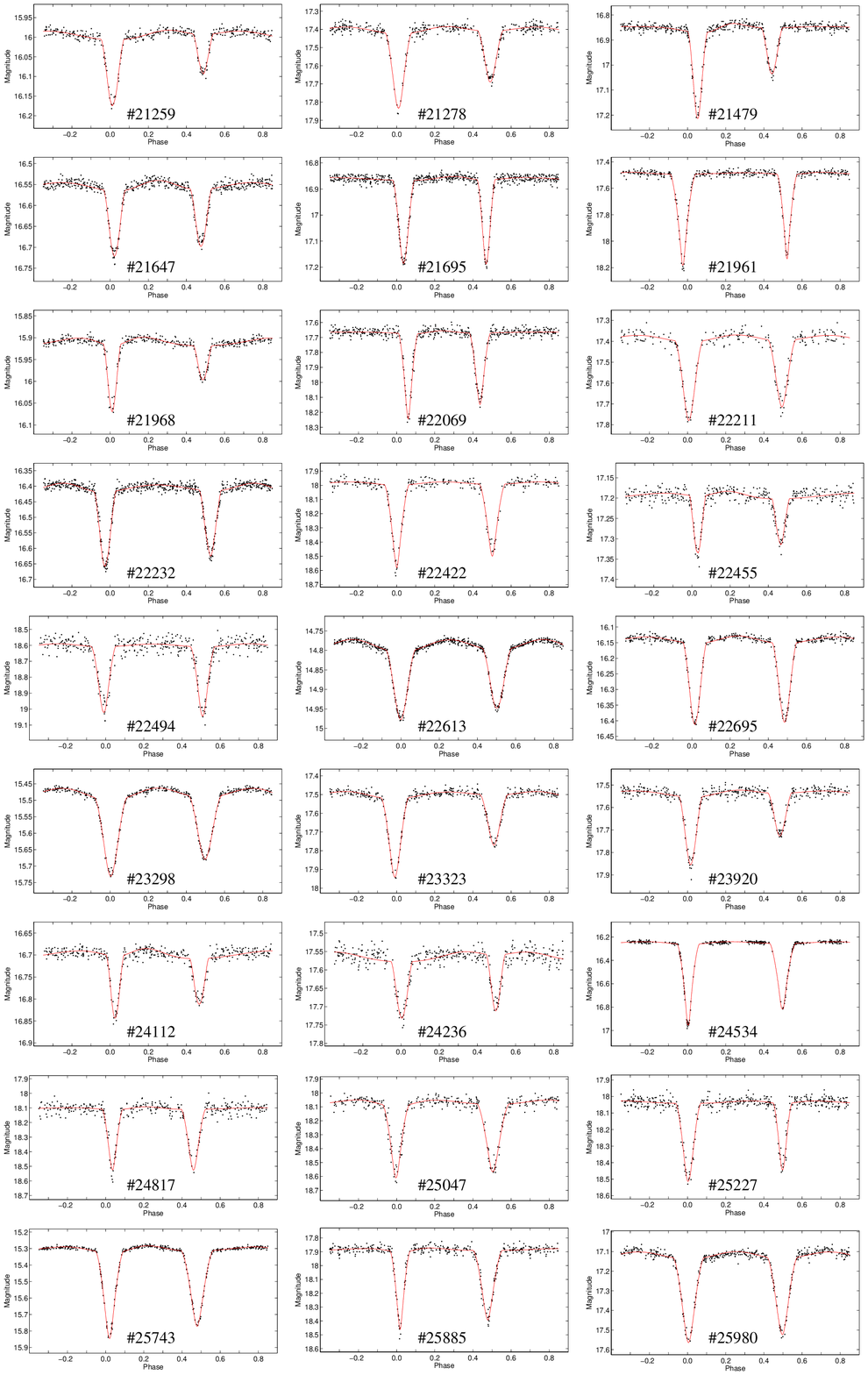}
  \caption{Plot of the light curves, part 6.}
  \label{FigLC6}
\end{figure*}

\begin{figure*}
  \centering
  \includegraphics[width=0.82\textwidth]{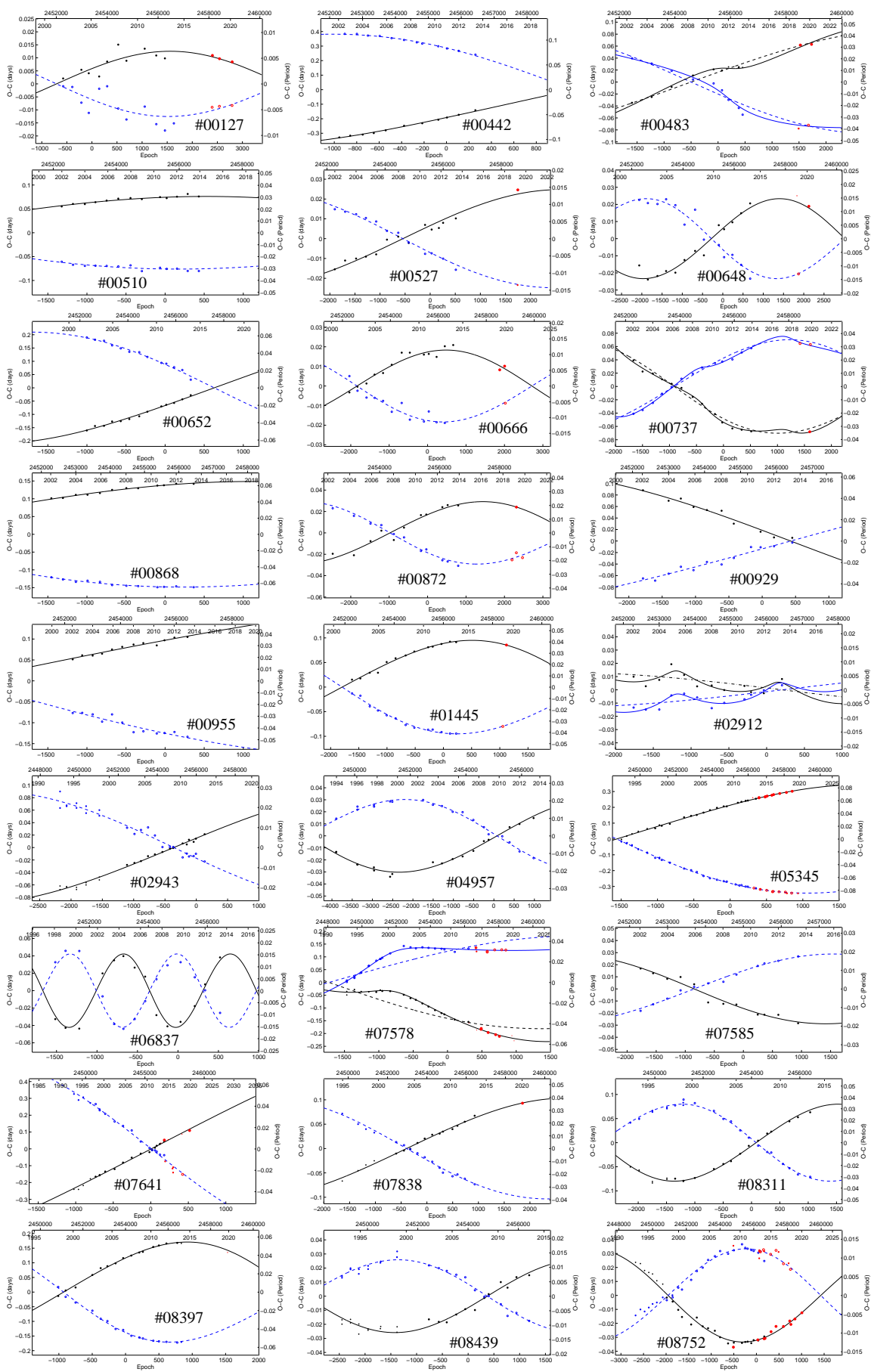}
  \caption{Plot of the $O-C$ diagrams, part 1. Final fits of the apsidal motion are
  plotted as black and blue curves (for primary and secondary minima), while
  the individual points represent the times of eclipses. The larger
  the symbol, the higher the weight (which was assigned according to its uncertainty).
  New observations with DK154 are shown in red. In cases where an additional third body
  was found, the final fit is plotted (solid lines) together with only the apsidal
  motion fits (dashed lines).}
  \label{FigOC1}
\end{figure*}

\begin{figure*}
  \centering
  \includegraphics[width=0.845\textwidth]{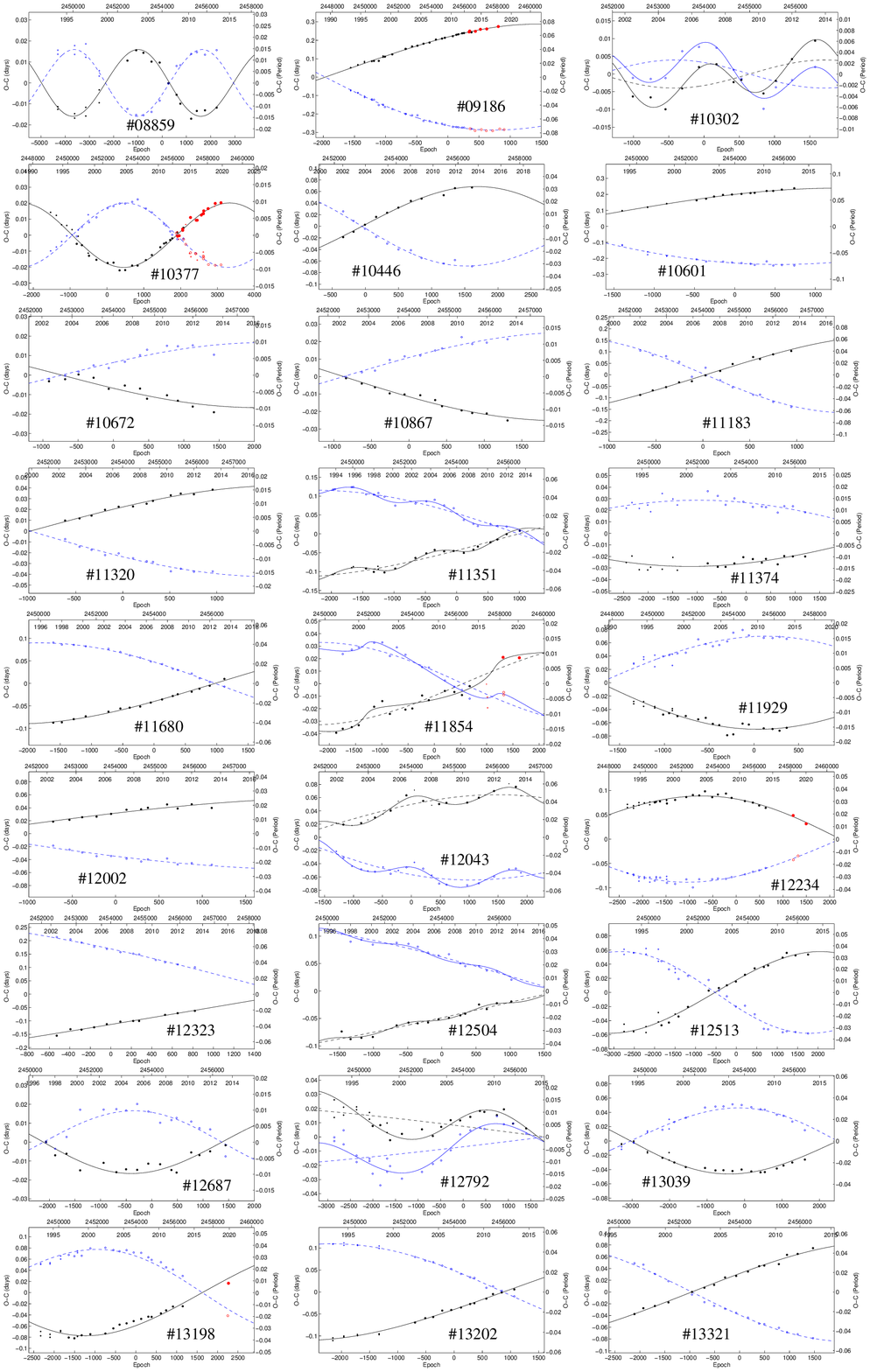}
  \caption{Plot of the $O-C$ diagrams, part 2.}
  \label{FigOC2}
\end{figure*}

\begin{figure*}
  \centering
  \includegraphics[width=0.845\textwidth]{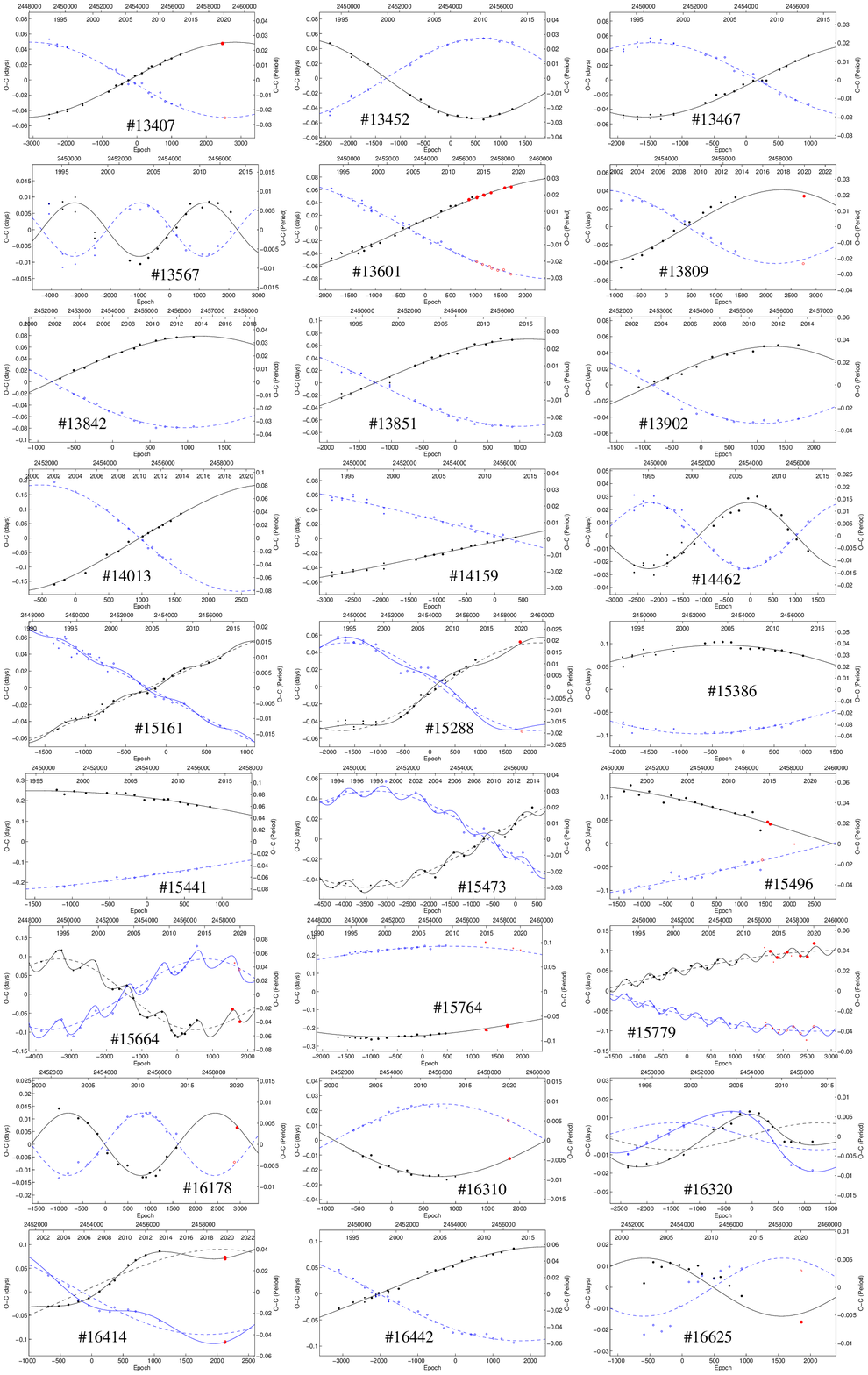}
  \caption{Plot of the $O-C$ diagrams, part 3.}
  \label{FigOC3}
\end{figure*}

\begin{figure*}
  \centering
  \includegraphics[width=0.845\textwidth]{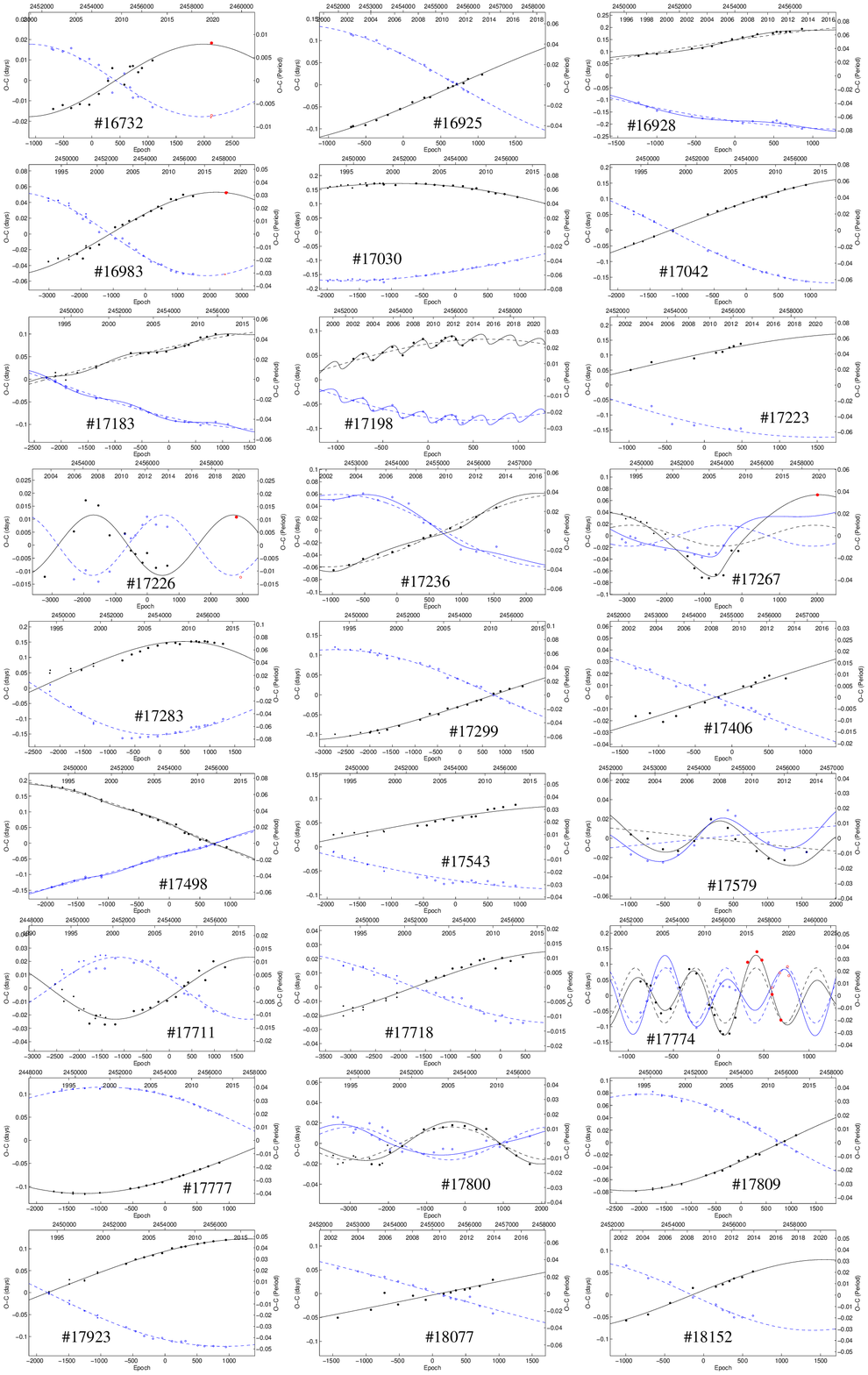}
  \caption{Plot of the $O-C$ diagrams, part 4.}
  \label{FigOC4}
\end{figure*}

\begin{figure*}
  \centering
  \includegraphics[width=0.845\textwidth]{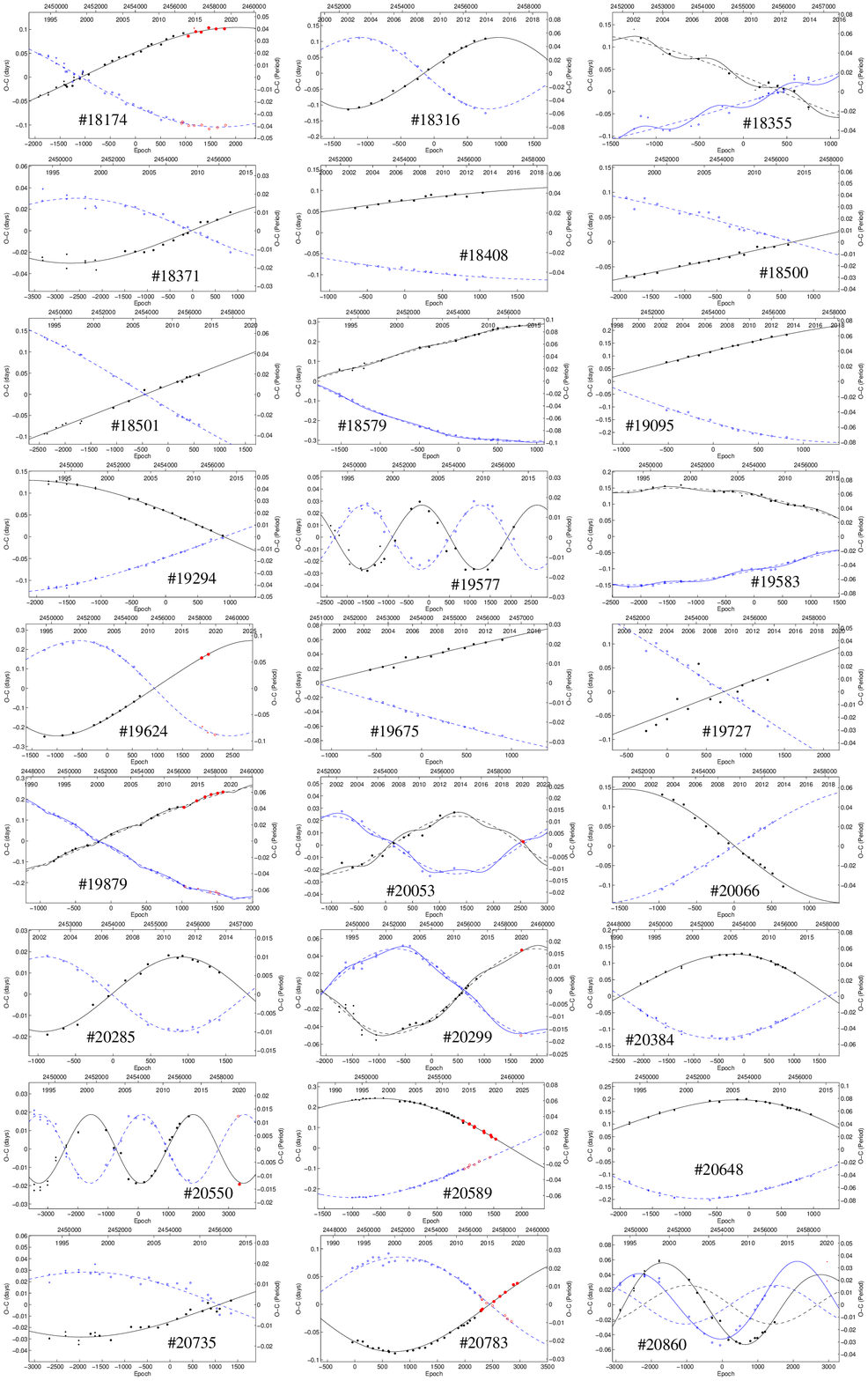}
  \caption{Plot of the $O-C$ diagrams, part 5.}
  \label{FigOC5}
\end{figure*}

\begin{figure*}
  \centering
  \includegraphics[width=0.845\textwidth]{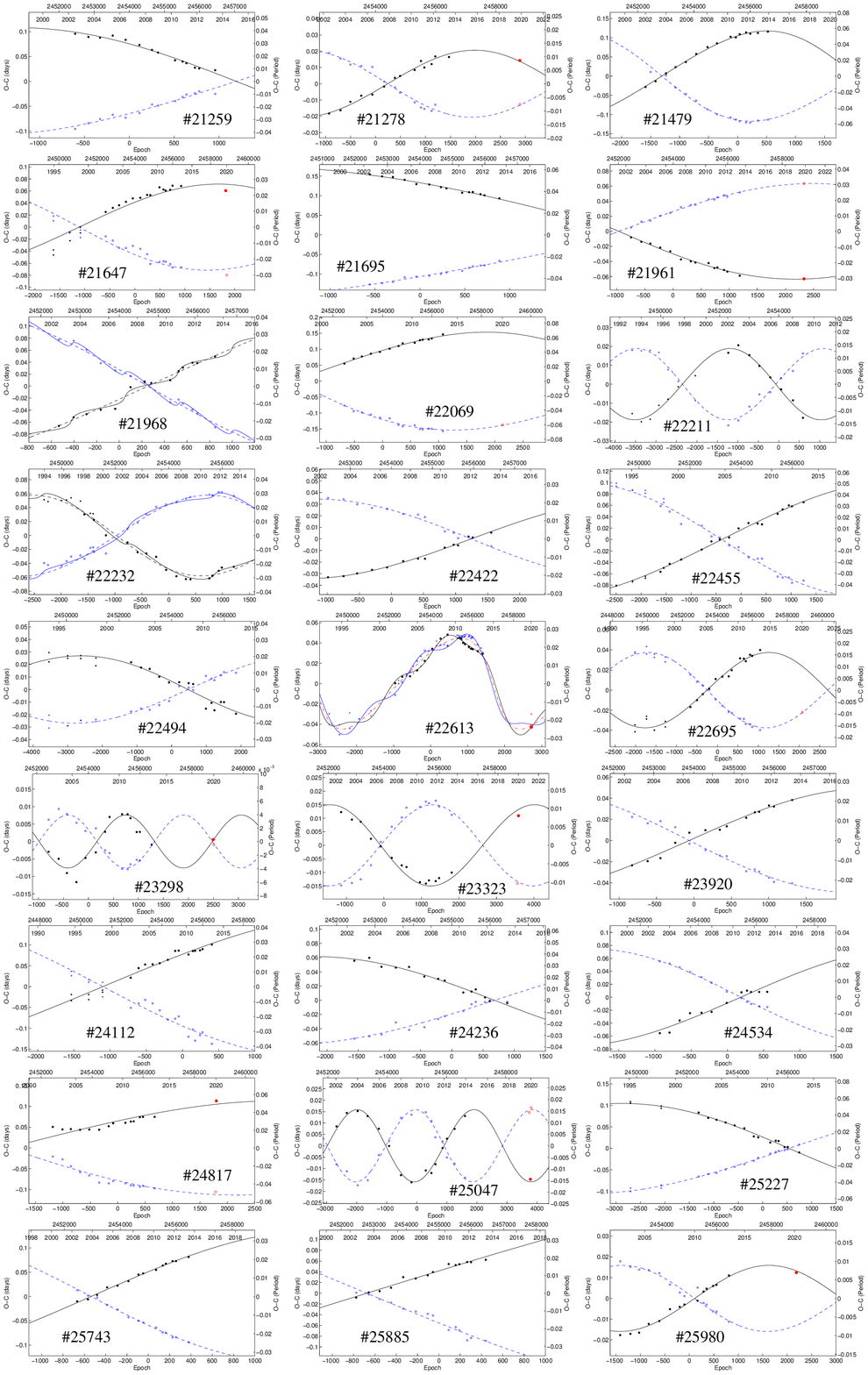}
  \caption{Plot of the $O-C$ diagrams, part 6.}
  \label{FigOC6}
\end{figure*}

\begin{figure*}
  \includegraphics[width=0.95\textwidth]{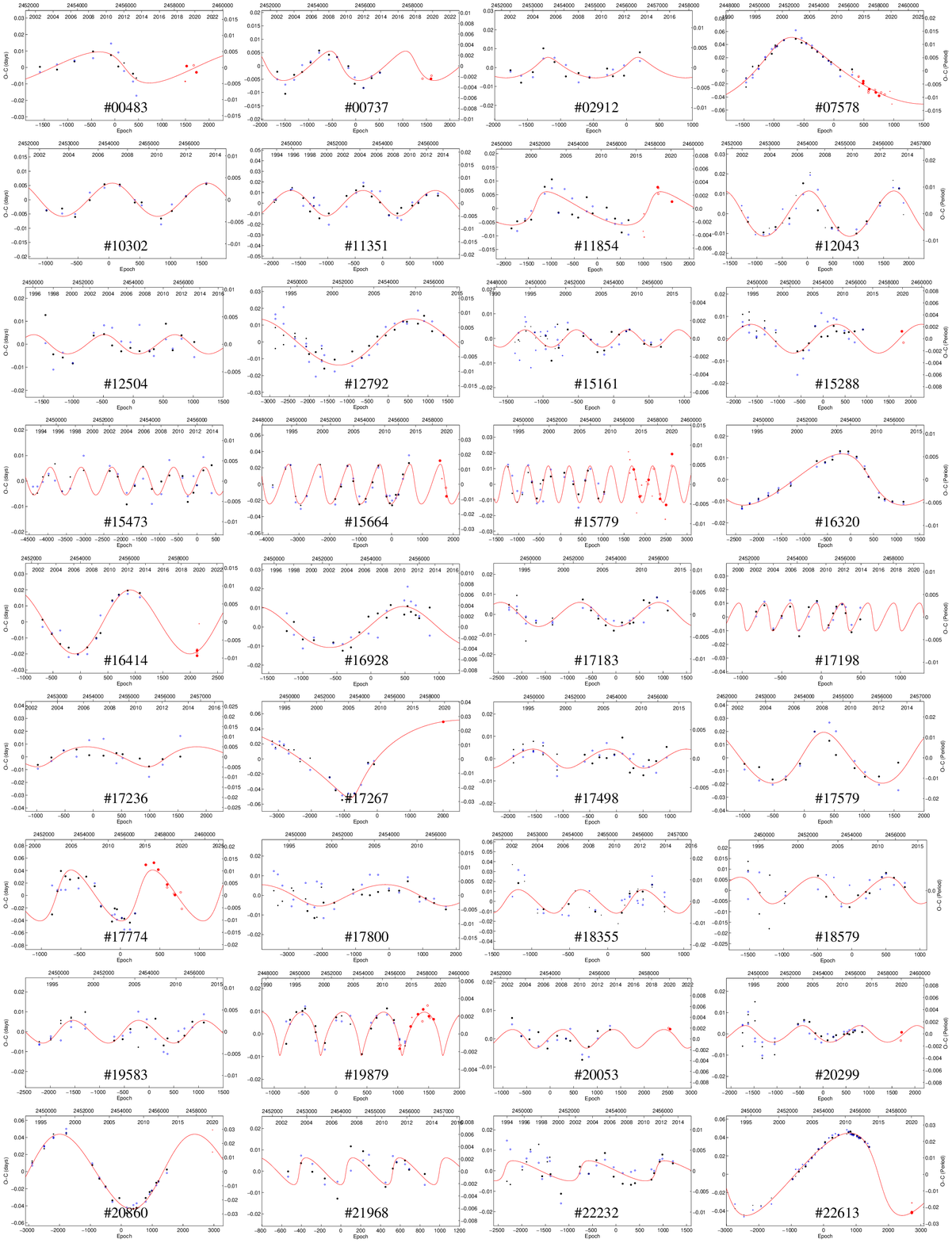}
  \caption{Plot of the $O-C$ diagrams after subtraction of apsidal motion fits. The LTE variations are plotted as red curves. }
  \label{FigOClite}
\end{figure*}

\section{Results}  \label{results}

Due to the very large number of individual systems, we cannot focus on each of the stars analysed
in our sample. We would thus like to focus only on the most interesting results and systems that
deserve our attention, and to encourage  future observers and investigators to consider these
particular systems.

All of our results are divided into several parts. First, the classical LC modelling and its
results are given below in Tables \ref{LCOCparam} and \ref{LCOCparam2} and the plots in Figs.
\ref{FigLC1} -- \ref{FigLC6}.  Second, the apsidal motion parameters are given in Tables
\ref{LCOCparam} and \ref{LCOCparam2}, while the fits can be seen in Figs. \ref{FigOC1} --
\ref{FigOC6}. These figures show the apparent period variations of the particular stars via the
eclipse timing variation together with the fits. Finally, the systems where some additional
variations in the times of eclipses were detected and showed unambiguous period variability were
additionally fitted assuming the light-time effect hypothesis \citep{Mayer1990}. This hypothesis
assumes a third body orbiting the inner eccentric eclipsing binary around a common barycentre.
However, the potential detectability of any such body is highly dependent on the third body's mass,
its orbit, orientation, period, etc. Therefore, the number of systems detected with our method
should be considered  a lower limit, and any definite conclusions about the multiplicity fraction
are still rather speculative  because the incompleteness fraction would be rather high (see
discussion   in Section \ref{discussion}). Altogether 36 such systems were found for which the
residuals after the apsidal motion fits were successfully fitted with the light-time effect
hypothesis.  All of these are shown in Fig. \ref{FigOClite}, while five parameters of these fits
are given in Table \ref{LITEparam}.

At this point it would be useful to recall some basic equations of the light-time effect and also
our analysis approach. The observed deviation of individual minima in the $O-C$ diagram can be
described by
 \begin{equation} \label{OCefem} (O-C) = JD_i - HJD_0 - P \cdot E - (O-C)_{LTE} - (O-C)_{AM},\end{equation}
where the particular minimum $JD_i$ at epoch $E$ is being compared with its theoretical prediction
also taking into account   the light-time effect term $(O-C)_{LTE}$ and the apsidal motion term
$(O-C)_{AM}$. The light-time effect can be expressed as
\begin{equation} \label{OClite} (O-C)_{LTE} = \frac{A} {\sqrt{1-e_3^2\cos^2\omega_{3}}} \left[ {{ \frac{(1-e_3^2)\sin(\nu + \omega_{3})}{1+e_3
\cos\nu}+ e_3 \sin\omega_{3}}} \right], \end{equation}
 where $A$ is the semi-amplitude,
$e_3$ is the eccentricity of the third body orbit, $\omega_3$ its argument of periastron, and $\nu$
its true anomaly. The mass function of this third-body orbit can be expressed as
  \begin{equation} \label{fm3} f(m_3) = \frac {(m_3 \sin i)^3} {(m_1+m_2+m_3)^2} = \frac{1}{P_3^2}
  \cdot {\left[ \frac{173.15 \cdot A} {\sqrt{1-e_3^2\cos^2\omega_3}} \right]^3}, \end{equation}
where $i$ stands for the inclination between the mutual orbit and the plane of the sky, and $m_1,
m_2, m_3$ are the masses of the primary, secondary, and tertiary components, respectively. For the
second expression it is necessary to input the units of the outer period $P_3$ in years and the
semi-amplitude $A$ in days. With this mass function we can at least make some preliminary estimates
about the third body's mass. However, for some more reliable derivation precise masses and also the
inclination are needed. We also note that in a triple system, there is in principle a much more
complicated influence of the third perturbing body on the apsidal motion of the eccentric inner
pair (see e.g. \citealt{2015MNRAS.448..946B}, or \citealt{2019EAS....82...99B}) in addition to  the
classical tidal and the relativistic contributions.

The total ETV component due to the action of a third body is composed in principle of several
contributions. Here we assume only the most dominant one, the classical geometrical light-time
effect term. However, in general there are also other dynamical terms. In principle, these effects
can be large or even dominate over the classical light-time effect. In our sample, where we deal
with quite short orbital periods of the inner pairs and quite long periods of the outer bodies, we
can state that for the majority of systems the light-time effect is certainly the dominant one. In
\cite{2013ApJ...768...33R}   a   physical delay due to the third body is introduced, which has its
amplitude proportional to the ratio ($P^2/P_3$). Hence, for majority of our systems third-body
period that is too long makes the amplitude of this physical delay too small. On the other hand,
there are also
 other longer-period perturbations (see e.g. \citealt{2013ApJ...768...33R},   Section 4.2),
which have typical periods of the order of $P_3^2/P$, and that cause for example the orbital
precession and change in the eclipse depths. In our sample the timescales of the longer-period
perturbations are typically longer than a human lifetime. Both these ratios $P^2/P_3$, and
$P_3^2/P$ are given in the last columns of Table \ref{LITEparam}.

The question arises of  whether these systems show some concordance with the systems with a
detected third light contribution. However, we note that this link is not so unambiguous. For some
of the potential triple systems the third bodies should have only negligibly low luminosities. On
the other hand, for some systems, the detection of a third light could also be explained by an
unbound star along the same line of sight. This situation is nothing novel in these very dense
stellar fields in the LMC. However, mostly the detection of the light-time effect agrees with the
non-zero value of the third light for a particular system. A typical example would be   OGLE
LMC-ECL-15664, which was identified as a system showing the  largest mass function, as well as the
largest value of the third light in the LC solution.

The quality of the apsidal motion fits in Figs. \ref{FigOC1} -- \ref{FigOC6} is mostly reasonable,
but strongly depends on the value of eccentricity of the orbit (i.e. the amplitude of apsidal
motion in the $O-C$ diagram), and its apsidal motion rate (i.e. the period of apsidal advance,
because longer periods only have  small parts covered with data). The quality of the individual
times of eclipses is  comparable for different systems (usually of the order of 0.001-0.0001 d),
only  those with shallow eclipses and/or longer orbital periods have less precise times of eclipse.

The most interesting definitely seem to be those systems with third bodies and with a very rapid
apsidal motion. Stars that have very fast apsidal motion usually have very short orbital periods.
Among these, there was discovered a record-breaking system,  OGLE LMC-ECL-17226, which has an
eccentric orbit despite its short period (P = 0.9879314 day). This is, as far as we know, the
shortest eccentric main sequence star ever discovered (the previous record of about 1.016 days
belongs to the star V456~Oph located in our Galaxy; see \citealt{2017AcA....67..257W}). The apsidal
motion of OGLE LMC-ECL-17226 is also rather fast (about 12 yr), while its eccentricity is very low
(about 0.037).

Among other stars (in addition to OGLE LMC-ECL-17226), additional systems having a very rapid
apsidal motion of about ten years or even faster were also found: OGLE LMC-ECL-06837 (U = 9.9\;yr),
$\#$17774 (U = 7.2\;yr), -22613 (U = 6.6\;yr), -23298 (U = 12.2\;yr), -25047 (U = 10.8\;yr). To
date, the system with the fastest apsidal advance in our Galaxy is V490~Cyg
\citep{2011A&A...527A..43Z} having U = 18.6\;yr, while out of our Galaxy, it is the star from the
SMC named OGLE SMC-ECL-2194 having U = 7.1\;yr \citep{2015AJ....150....1H}. For the list of all our
analysed systems and their results, see Tables \ref{LCOCparam} and \ref{LCOCparam2}. It is
important to note that the ephemerides given in Tables \ref{LCOCparam} and \ref{LCOCparam2} are not
always suitable for planning future observations. They represent only the value $O-C=0$ in the
diagrams in Figs. \ref{FigOC1}-\ref{FigOC6}. Hence for eclipse predictions, corrections are needed,
especially in   cases where the eccentricity is high and the apsidal motion amplitude significant.

Another interesting group of stars is the one where the third-body contribution to ETV caused by
the light-time effect is even larger than the contribution from the apsidal motion itself. These
are: OGLE LMC-ECL-10302, -16320, -17267, -20860, and -22613.  The dominance of the light-time
effect can be caused by two different phenomena: a very massive third body causing large amplitude
of ETV, or in some systems only very low eccentricity of the inner orbit. In our cases shown here,
both of these explanations play a role together.

\begin{figure}
  \includegraphics[width=0.49\textwidth]{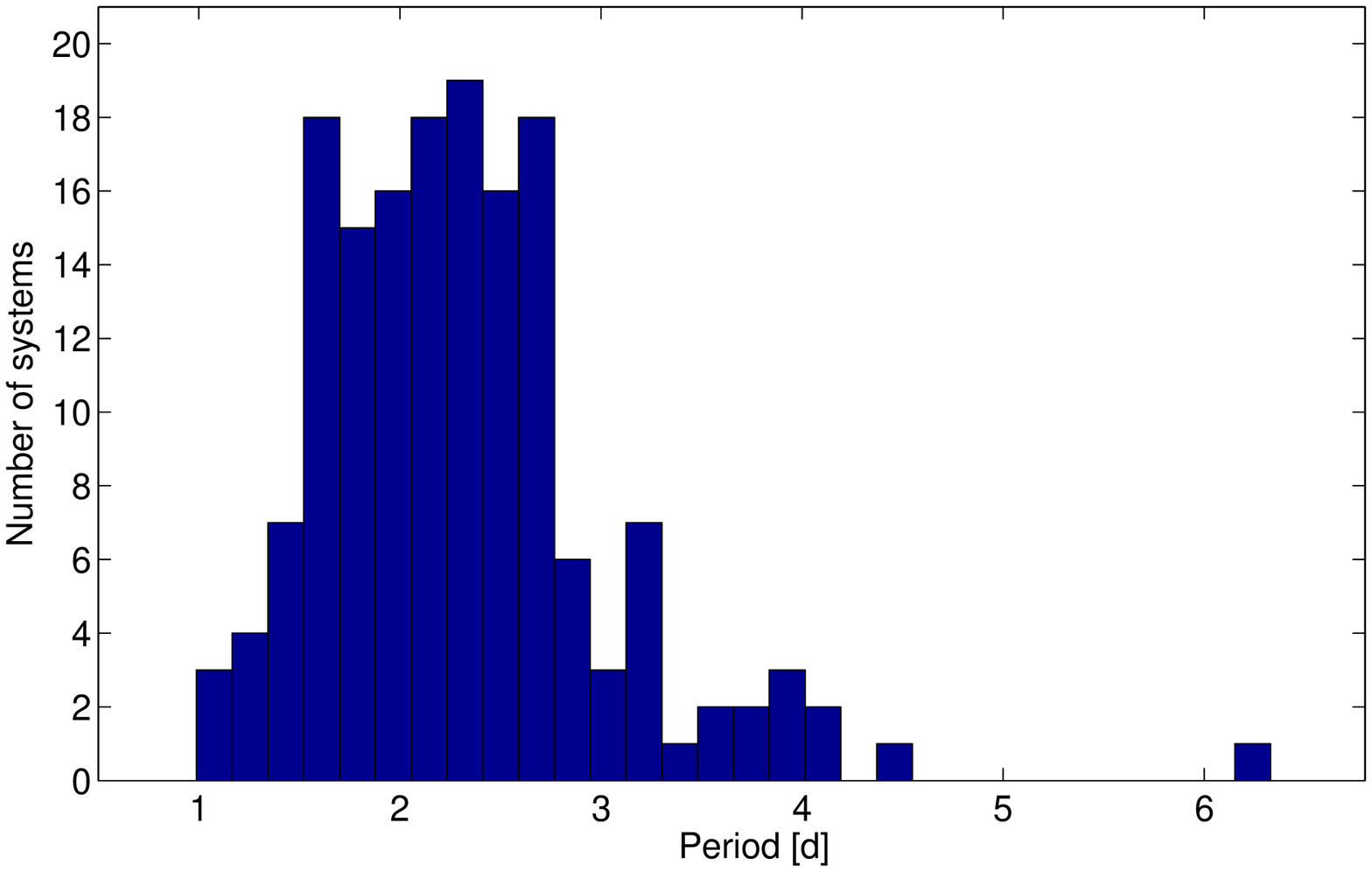}
  \caption{Histogram of orbital periods of the analysed systems.}
  \label{FigPerhistogram}
\end{figure}

\section{Properties of the sample}

First we   look at the distribution of the orbital periods of the stars in our sample. Obviously,
these are influenced by a strong selection effect, mainly due to our method of selection. Only
those systems having an adequately short orbital period (usually $P\lesssim3$d) can be studied
using our methods (and having enough data for such an analysis). For this reason the plot in Fig.
\!\ref{FigPerhistogram} shows a vast majority of systems from 1 to 3 days. Shorter-period systems
are usually circular, while the longer-period ones usually have too slow apsidal motion to show a
significant change over two decades, which is the typical time span of our data.

\begin{figure}
  \includegraphics[width=0.49\textwidth]{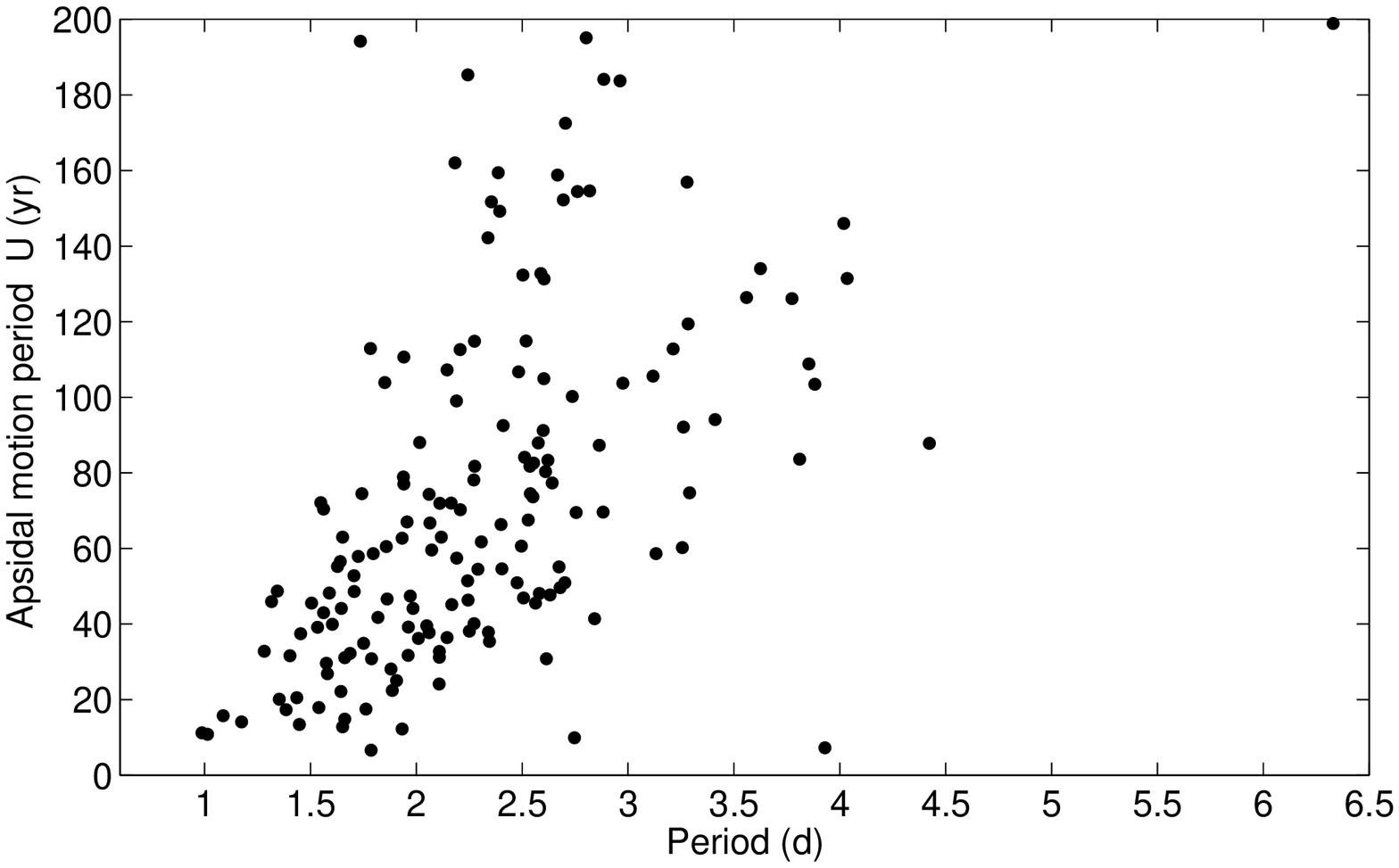}
  \caption{Diagram showing apsidal motion period vs orbital period.}
  \label{FigPer_U}
\end{figure}

Another diagram in Fig. \!\ref{FigPer_U} shows how the apsidal motion period depends on the orbital
period of the binary. Longer-period  systems usually also have  slower apsidal motion, while the
fastest apsidal motion should preferably be detected for shorter orbital periods.

\begin{figure}
  \includegraphics[width=0.49\textwidth]{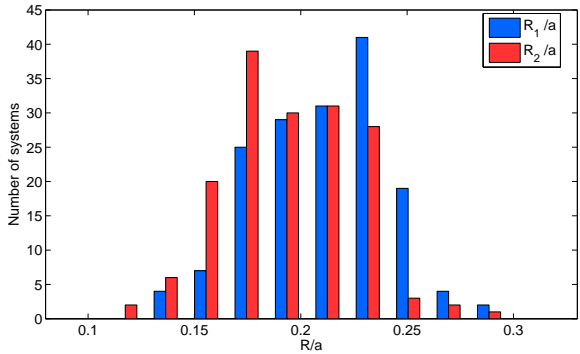}
  \caption{Histogram of relative radii of all systems for both eclipsing components.}
  \label{FigR12histogram}
\end{figure}

In Fig. \!\ref{FigR12histogram} we show the distribution of the relative radii  resulting from our
light curve analysis (for the primary and secondary components). It is clearly seen that
preferentially those systems having relative radii between 0.15 and 0.25 were usually studied. This
is again a selection effect of our method. Systems having too distant components usually have too
slow apsidal motion ($>$ 200 yr) to be detectable with our approach, and systems closer to each
other were already circularised.

\begin{figure}
  \includegraphics[width=0.49\textwidth]{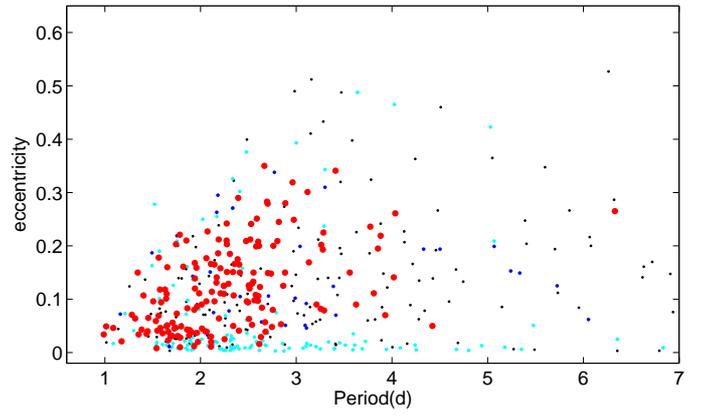}
  \caption{Plot showing the distribution of eccentricities vs orbital periods for eclipsing binaries. The red dots are our new data from the present paper,
  which are being compared with the other LMC/SMC systems. The blue dots are taken from our previous analyses, and the cyan dots are taken from publications by Hong et al.
  (see Table in Section \ref{intro} for reference). A comparison with galactic eclipsing binaries as published in the catalogue of \cite{2018ApJS..235...41K} are plotted
  in black. Circularisation in shorter periods is clearly visible here.}
  \label{FigEccPer}
\end{figure}

In Fig. \ref{FigEccPer} we plotted the eccentricity versus orbital period. For comparison we also
added other compilations and apsidal motion studies from LMC/SMC from Table \ref{LMCSMCEEBs}, and
the apsidal motion systems  published in the catalogue of \cite{2018ApJS..235...41K}, who studied
galactic systems. It is clearly seen that our present sample incorporates mainly the shorter-period
systems (mostly below 4 days), but the overall tendency to detect the higher eccentricities only
with longer periods is obvious. Hence, the circularisation in more compact systems is also much
faster  in the LMC galaxy (see e.g. \citealt{1988ApJ...324L..71T}).

\begin{figure*}
  \centering
  \includegraphics[width=1.05\textwidth]{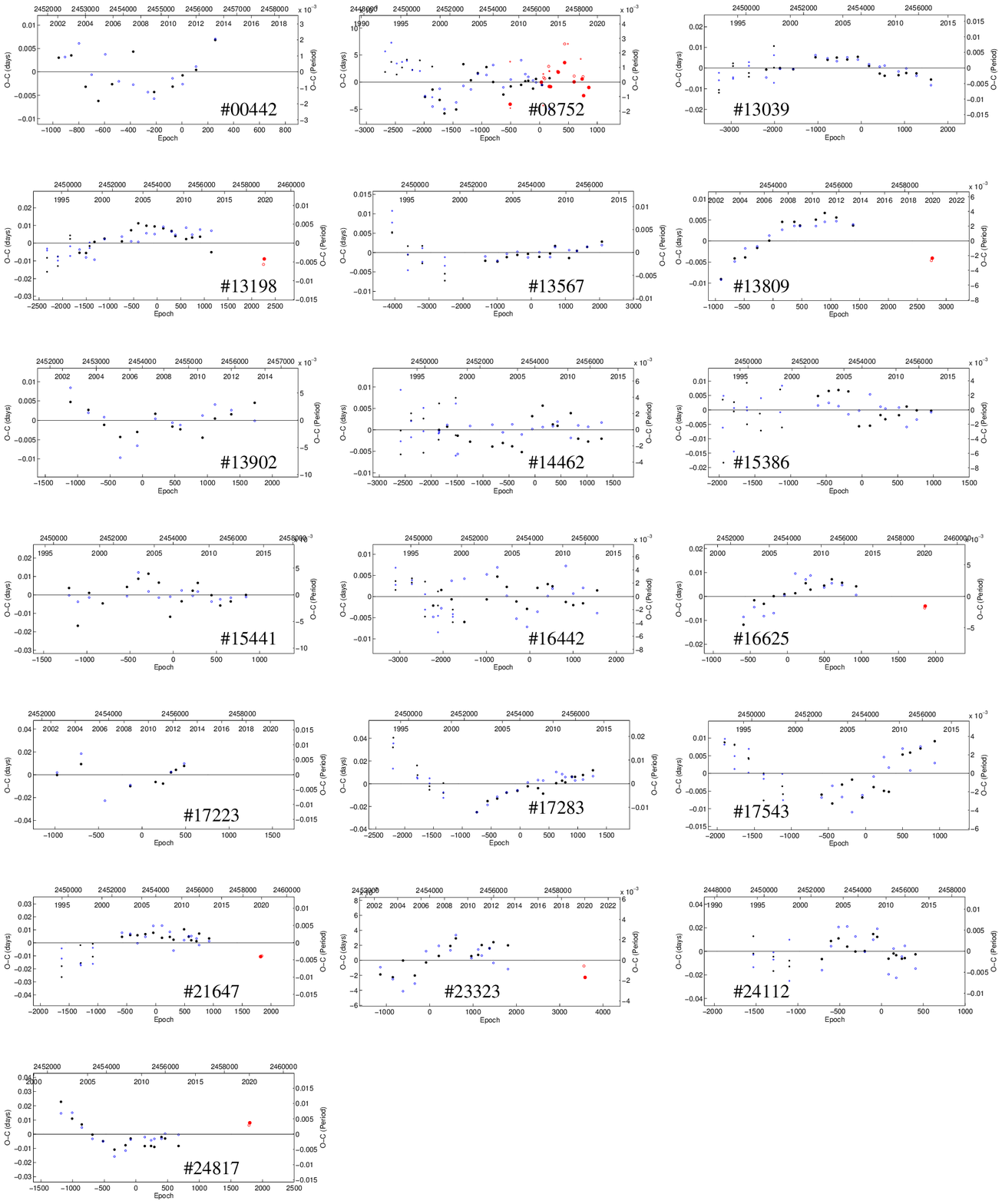}
  \caption{Plot of the $O-C$ diagrams after subtraction of apsidal motion fits for cases where there appears to be some
  variation of residuals, but it is still rather questionable. More data are needed.}
  \label{FigOCresid}
\end{figure*}

 \section{Discussion and conclusions}  \label{discussion}

Our new analysis of 162 eccentric eclipsing binaries in LMC presents the largest sample studied to
date in one paper (both of the LC and apsidal motion) and provide us with  unique material for
comparison with other results.

Because we have limited information about the stars, and no spectroscopy for these targets, the
inferred parameters are only relative ones (like the relative radii), and the masses are completely
unknown. The uncertainties of masses inferred from the photometric indices are too large, hence
they cannot easily be used to derive  the internal structure constants. Only future spectroscopic
monitoring would help us in this aspect.

The ETVs due to the light-time effect term are usually rather small, which also yielded small mass
function values (with only a few exceptions). The other ETV effects are  negligible, or would be
very slow (see the last columns of Table \ref{LITEparam}). Hence, for instance we cannot hope to
find any dynamical indication of these third bodies, for example via changing the inclination angle
through eclipse depth variations (only one system has the typical period of this nodal regression
below 1000~years).

On the other hand, the presented study of residuals and fitting of the light-time effect (see Figs.
\ref{FigOClite}) is only one part of the whole sample. Other systems were found to show additional
variation, but they are still not well-covered, have too long periods, or too low amplitudes. For
these reasons the variation is visible, but for a proper analysis we need more data spanning  a
longer time interval or  better data to reduce the scatter. This is why we only briefly mention
these systems in Fig. \ref{FigOCresid}, where the residuals after the subtraction of the apsidal
motion are plotted, but no light-time effect fit is given yet. This should encourage observers to
collect new data for these systems to prove or reject the hypothesis of the third body.

If we take into account   the proved 36 systems with the light-time effect fits in  Figs.
\ref{FigOClite} and add the 19 suspected ones from Fig. \ref{FigOCresid}, then the total fraction
of potential triples in our sample grows from 20\% to about 34\%. We are aware that our
preprocessing and selection of suitable systems for this analysis can provide some bias towards
these potential triples. However, as was already stated in Sect. \ref{methods}, for most of the
discarded systems the criterion was the slow rate of apsidal motion. This  is probably not directly
connected with the multiplicity, hence its role should be small. However, what should play a role
is that the stars with very small photometric amplitude (shallow eclipses) were also neglected.
These systems might also  be triples because the high value of the third light makes the eclipses
shallower. Hence, we would probably miss these potential triples rather than overestimate their
fraction in the LMC eclipsing sample. Such a large fraction of potential triples is several times
more than what was found in their sample by \cite{2018ApJS..235...41K}. However, they studied the
stars in our Galaxy, where the field stars are of much later spectral type; hence, the multiplicity
fraction is much lower \citep{2013ARA&A..51..269D}. Therefore, our number should only be compared
with the suggested number of multiples in the early-type population in LMC. According to
\cite{2013ARA&A..51..269D}, the probability of a star to be a multiple for these massive stars
should be very high and reach almost 100\% for the O and B stars. Hence, our rough estimation of
1/3 should be considered a lower limit due to the methodology incompleteness limitations (e.g. long
periods and short periods cannot be detected with our method; masses and inclinations play a
crucial role for a light-time effect to emerge in our residuals).

\begin{acknowledgements}
We are thankful to an anonymous referee for the critical suggestions greatly improving the quality
of the whole manuscript. This work was supported by the grant MSMT INGO II LG15010, grant
GA\,\v{C}R 18-05665S, and grant Primus/SCI/17. Marek Skarka acknowledges the OP VVV project
Postdoc@MUNI (No. CZ.02.2.69/0.0/0.0/16027/0008360). We would like to thank J.Li\v{s}ka, K.Hornoch,
L.Pilar\v{c}\'{\i}k, and J.Vra\v{s}til for obtaining some of the photometric data. We also do thank
the OGLE and MACHO teams for making all of the observations easily public available. We are also
grateful to the ESO team at the La Silla Observatory for their help in maintaining and operating
the Danish telescope. This research has made use of the SIMBAD and VIZIER databases, operated at
CDS, Strasbourg, France and of NASA Astrophysics Data System Bibliographic Services.
\end{acknowledgements}

 \begin{appendix}
 \section{Appendix}\label{apendix}

\begin{table*}
\caption{Relevant information for the analysed systems.}  \label{InfoSystems}
 \scriptsize
  \centering \scalebox{0.97}{
\begin{tabular}{lccccccccccc}
   \hline\hline\noalign{\smallskip}
 \multicolumn{3}{c}{S\,y\,s\,t\,e\,m\,\,\,\, n\,a\,m\,e}     &     RA      &      DE     & $V_{\rm max}^{\,A}$ &  $(B-V)^{B}$ & $(B-V)_0^{B}$ & $(B-V)^{C}$ & $(B-V)_0^{C}$ \\
 \multicolumn{1}{c}{OGLE III}    & OGLE II      & MACHO      &             &             & [mag]               &    [mag]     &   [mag]       &   [mag]     & [mag]         \\
    \hline\noalign{\smallskip}
 \object{OGLE LMC-ECL-00127}     &              &            & 04:37:08.25 & -69:06:55.1 & 17.60 &    -- &    -- &  -0.021& -0.097   \\ 
 \object{OGLE LMC-ECL-00442}     &              &            & 04:43:14.95 & -70:33:58.6 & 15.45 &    -- &    -- &  -0.116& -0.295   \\ 
 \object{OGLE LMC-ECL-00483}     &              &            & 04:43:38.64 & -69:43:34.6 & 18.58 &    -- &    -- &   0.153& -0.153   \\ 
 \object{OGLE LMC-ECL-00510}     &              &            & 04:43:51.87 & -68:49:32.0 & 17.56 &    -- &    -- &  -0.458&  0.139   \\ 
 \object{OGLE LMC-ECL-00527}     &              &            & 04:44:02.05 & -67:20:33.2 & 18.07 &    -- &    -- &  -0.308&  0.151   \\ 
 \object{OGLE LMC-ECL-00648}     &              &            & 04:45:18.51 & -69:43:02.7 & 17.75 &    -- &    -- &   0.001& -0.160   \\ 
 \object{OGLE LMC-ECL-00652}     &              &            & 04:45:23.73 & -69:09:55.7 & 16.61 & -0.05 &-0.217 &  -0.070& -0.181   \\ 
 \object{OGLE LMC-ECL-00666}     &              &            & 04:45:35.21 & -68:29:14.3 & 17.38 &    -- &    -- &   0.061& -0.299   \\ 
 \object{OGLE LMC-ECL-00737}     &              &            & 04:46:17.60 & -67:53:06.5 & 17.05 &    -- &    -- &  -0.052& -0.220   \\ 
 \object{OGLE LMC-ECL-00868}     &              &44.1138.118 & 04:47:23.92 & -69:38:28.9 & 17.92 &  0.19 &-0.073 &   0.007& -0.139   \\ 
 \object{OGLE LMC-ECL-00872}     &              &45.1157.218 & 04:47:25.72 & -68:25:12.4 & 18.54 &    -- &    -- &   0.135& -0.189   \\ 
 \object{OGLE LMC-ECL-00929}     &              &            & 04:47:52.41 & -66:52:28.8 & 18.47 &    -- &    -- &  -0.033&  0.151   \\ 
 \object{OGLE LMC-ECL-00955}     &              &            & 04:48:01.92 & -70:22:44.3 & 18.61 &    -- &    -- &   0.126& -0.131   \\ 
 \object{OGLE LMC-ECL-01445}     &              & 44.1626.46 & 04:50:35.32 & -69:25:20.2 & 16.82 &  0.13 &-0.108 &  -0.014& -0.229   \\ 
 \object{OGLE LMC-ECL-02912$^*$} &              &            & 04:55:14.81 & -68:56:18.5 & 17.82 & -0.26 & 0.034 &   0.267& -0.154   \\ 
 \object{OGLE LMC-ECL-02943}     &              &18.2360.151 & 04:55:20.88 & -68:51:10.1 & 17.62 &    -- &    -- &   0.031& -0.066   \\ 
 \object{OGLE LMC-ECL-04957}     &              & 24.3223.52 & 05:00:32.62 & -67:48:17.1 & 18.08 &    -- &    -- &   0.251& -0.026   \\ 
 \object{OGLE LMC-ECL-05345}     &  SC15 152403 &18.3449.3635& 05:01:40.18 & -68:51:05.5 & 15.38 & -0.06 &-0.271 &  -0.040& -0.274   \\ 
 \object{OGLE LMC-ECL-06837}     &  SC13 66307  &19.4063.100 & 05:05:07.12 & -68:16:22.9 & 17.17 &    -- &    -- &   0.056& -0.231   \\ 
 \object{OGLE LMC-ECL-07578}     &  SC13 183860 &19.4300.349 & 05:06:43.25 & -68:36:11.4 & 15.76 &    -- &    -- &  -0.072& -0.220   \\ 
 \object{OGLE LMC-ECL-07585$^*$} &              &25.4321.184 & 05:06:44.73 & -67:12:41.8 & 18.40 &    -- &    -- &   0.502& -0.340   \\ 
 \object{OGLE LMC-ECL-07641}     &  SC13 194100 &19.4302.319 & 05:06:51.94 & -68:25:46.5 & 14.22 &    -- &    -- &  -0.184& -0.195   \\ 
 \object{OGLE LMC-ECL-07838}     &              &52.4447.1225& 05:07:17.83 & -66:50:37.6 & 16.81 &    -- &    -- &   0.030& -0.174   \\ 
 \object{OGLE LMC-ECL-08311}     &  SC11 103746 & 1.4530.37  & 05:08:27.80 & -69:25:28.7 & 15.82 &  0.15 &-0.346 &  -0.111& -0.238   \\ 
 \object{OGLE LMC-ECL-08397$^*$} &  SC11 162262 &            & 05:08:38.89 & -68:45:45.7 & 14.69 &    -- &    -- &   0.085& -0.531   \\ 
 \object{OGLE LMC-ECL-08439}     &              &19.4663.3317& 05:08:44.93 & -68:34:26.3 & 16.97 &    -- &    -- &   0.045& -0.170   \\ 
 \object{OGLE LMC-ECL-08752}     &  SC11 331573 &79.4779.34  & 05:09:29.21 & -68:55:02.6 & 14.92 & -0.07 &-0.302 &  -0.238& -0.117   \\ 
 \object{OGLE LMC-ECL-08859}     &              &20.4795.146 & 05:09:45.57 & -67:51:02.8 & 17.74 &    -- &    -- &  -0.019& -0.019   \\ 
 \object{OGLE LMC-ECL-09186}     &  SC10 16968  &5.4894.3904 & 05:10:28.64 & -69:20:47.9 & 15.73 &  0.18 &-0.257 &  -0.012& -0.229   \\ 
 \object{OGLE LMC-ECL-10302}     &              &            & 05:13:25.33 & -69:34:13.9 & 16.18 &  0.19 &-0.259 &     -- &    --    \\ 
 \object{OGLE LMC-ECL-10377}     &   SC9 121731 &5.5377.4656 & 05:13:36.22 & -69:22:41.4 & 16.25 &  0.32 &-0.382 &  -0.004& -0.216   \\ 
 \object{OGLE LMC-ECL-10446}     &              &            & 05:13:45.17 & -67:19:47.8 & 16.07 &   --  &    -- &  -0.269& -0.001   \\ 
 \object{OGLE LMC-ECL-10601}     &              &16.5523.48  & 05:14:08.60 & -67:41:58.7 & 16.84 &   --  &    -- &   0.000& -0.179   \\ 
 \object{OGLE LMC-ECL-10672}     &              &            & 05:14:18.68 & -66:37:26.7 & 18.03 &   --  &    -- &   0.019& -0.218   \\ 
 \object{OGLE LMC-ECL-10867}     &              &            & 05:14:49.59 & -67:20:32.8 & 17.06 &   --  &    -- &  -0.029& -0.197   \\ 
 \object{OGLE LMC-ECL-11183$^*$} &              &            & 05:15:33.20 & -70:20:21.0 & 17.33 &   --  &    -- &   0.840& -0.389   \\ 
 \object{OGLE LMC-ECL-11320}     &              &            & 05:15:51.42 & -67:42:41.2 & 17.07 &   --  &    -- &  -0.036& -0.138   \\ 
 \object{OGLE LMC-ECL-11351}     &   SC8 168297 &79.5746.319 & 05:15:57.02 & -68:58:04.6 & 17.58 &   --  &    -- &   0.178& -0.180   \\ 
 \object{OGLE LMC-ECL-11374}     &              &79.5750.262 & 05:16:01.13 & -68:45:31.9 & 18.35 &   --  &    -- &   0.090& -0.104   \\ 
 \object{OGLE LMC-ECL-11680}     &   SC8 199230 &            & 05:16:50.17 & -69:35:37.9 & 17.07 &  0.05 & 0.005 &    --  &    --    \\ 
 \object{OGLE LMC-ECL-11854}     &   SC8 312491 &78.5981.322 & 05:17:15.40 & -69:26:14.8 & 17.09 & -0.28 &-0.189 &  -0.082& -0.096   \\ 
 \object{OGLE LMC-ECL-11929$^*$} &   SC8 305721 &78.5981.107 & 05:17:29.97 & -69:27:58.5 & 15.83 & -0.27 &-0.168 &  -0.033& -0.097   \\ 
 \object{OGLE LMC-ECL-12002}     &              &            & 05:17:40.76 & -67:59:47.1 & 16.76 &   --  &    -- &   0.031& -0.218   \\ 
 \object{OGLE LMC-ECL-12043}     &              &            & 05:17:45.49 & -67:53:06.5 & 18.29 &   --  &    -- &   0.132& -0.142   \\ 
 \object{OGLE LMC-ECL-12234$^*$} &   SC7 150213 &78.6100.606 & 05:18:12.60 & -69:35:24.4 & 17.45 &   --  &    -- &    --  &    --    \\ 
 \object{OGLE LMC-ECL-12323}     &              &            & 05:18:25.01 & -70:24:08.9 & 18.03 &   --  &    -- &   0.274& -0.226   \\ 
 \object{OGLE LMC-ECL-12504$^*$} &   SC7 278663 &            & 05:18:48.41 & -69:33:57.5 & 17.52 &   --  &    -- &    --  &    --    \\ 
 \object{OGLE LMC-ECL-12513$^*$} &   SC7 278743 &78.6222.903 & 05:18:50.01 & -69:33:30.3 & 17.74 &  0.38 &-0.418 &    --  &    --    \\ 
 \object{OGLE LMC-ECL-12687}     &   SC7 287027 &            & 05:19:15.46 & -69:30:58.2 & 17.25 &   --  &    -- &   0.045& -0.042   \\ 
 \object{OGLE LMC-ECL-12792}     &   SC7 417231 &78.6344.776 & 05:19:31.41 & -69:26:59.9 & 18.33 &   --  &    -- &  -0.136&  0.124   \\ 
 \object{OGLE LMC-ECL-13039$^*$} &   SC7 397930 &78.6464.1043& 05:20:07.88 & -69:32:41.4 & 17.93 &   --  &    -- &   0.473& -0.387   \\ 
 \object{OGLE LMC-ECL-13198$^*$} &   SC6 58359  &78.6463.505 & 05:20:35.07 & -69:34:37.8 & 17.30 &   --  &    -- &    --  &    --    \\ 
 \object{OGLE LMC-ECL-13202}     &              & 3.6484.101 & 05:20:35.79 & -68:12:17.2 & 17.39 & -0.14 &-0.030 &   0.064& -0.215   \\ 
 \object{OGLE LMC-ECL-13321}     &   SC6 143503 &            & 05:20:52.65 & -69:52:24.3 & 18.04 &   --  &    -- &   0.407& -0.200   \\ 
 \object{OGLE LMC-ECL-13407}     &              & 3.6603.171 & 05:21:05.82 & -68:20:10.4 & 17.62 & 0.560 &-0.090 &  -0.011& -0.013   \\ 
 \object{OGLE LMC-ECL-13452}     &              & 16.6609.43 & 05:21:11.63 & -67:55:34.6 & 16.53 &   --  &    -- &  -0.074& -0.243   \\ 
 \object{OGLE LMC-ECL-13467}     &              & 80.6591.25 & 05:21:14.51 & -69:05:40.6 & 15.72 & -0.11 &-0.239 &   0.105& -0.277   \\ 
 \object{OGLE LMC-ECL-13567}     &              &80.6712.1405& 05:21:30.12 & -69:07:15.8 & 18.04 & -0.12 & 0.115 &   0.053&  0.007   \\ 
 \object{OGLE LMC-ECL-13601$^*$} &   SC6 322535 &78.6708.180 & 05:21:34.84 & -69:25:34.6 & 16.70 &   --  &    -- &   0.266& -0.371   \\ 
 \object{OGLE LMC-ECL-13809$^*$} &              &            & 05:22:02.86 & -67:58:46.6 & 16.93 &   --  &    -- &   0.154& -0.370   \\ 
 \object{OGLE LMC-ECL-13842}     &              &            & 05:22:07.88 & -67:48:56.2 & 16.22 &   --  &    -- &  -0.090& -0.275   \\ 
 \object{OGLE LMC-ECL-13851}     &              &3.6841.1394 & 05:22:08.65 & -68:37:15.4 & 16.97 &  0.15 &-0.176 &  -0.183& -0.104   \\ 
 \object{OGLE LMC-ECL-13902}     &              &            & 05:22:14.88 & -67:47:16.7 & 17.64 &   --  &    -- &  -0.096& -0.182   \\ 
 \object{OGLE LMC-ECL-14013}     &              &            & 05:22:28.93 & -68:04:54.9 & 17.19 &   --  &    -- &   0.090& -0.283   \\ 
 \object{OGLE LMC-ECL-14159}     &              &80.6835.204 & 05:22:46.12 & -69:00:50.6 & 17.45 &  0.43 &-0.170 &  -0.087& -0.098   \\ 
 \object{OGLE LMC-ECL-14462}     &   SC5 124709 &6.6940.6208 & 05:23:27.72 & -70:02:53.3 & 15.97 &   --  &    -- &  -0.137& -0.215   \\ 
 \object{OGLE LMC-ECL-15161}     &   SC5 349685 &77.7303.152 & 05:25:09.33 & -70:04:22.6 & 16.41 &   --  &    -- &   0.072& -0.119   \\ 
 \object{OGLE LMC-ECL-15288}     &              & 80.7316.43 & 05:25:29.32 & -69:10:56.0 & 15.97 & -0.10 &-0.218 &  -0.067& -0.189   \\ 
 \object{OGLE LMC-ECL-15386}     &              &80.7318.1393& 05:25:45.38 & -69:01:54.8 & 17.67 &   --  &    -- &  -0.091& -0.059   \\ 
 \object{OGLE LMC-ECL-15441}     &   SC4 220655 &            & 05:25:53.52 & -69:30:49.5 & 17.62 &   --  &    -- &   0.116& -0.133   \\ 
 \object{OGLE LMC-ECL-15473}     &              & 3.7443.37  & 05:25:57.49 & -68:46:36.1 & 17.10 &  0.08 &-0.099 &   0.155& -0.247   \\ 
 \object{OGLE LMC-ECL-15496}     &   SC4 177590 &            & 05:26:00.17 & -69:51:15.4 & 18.11 &   --  &    -- &  -0.200&  0.049   \\ 
 \object{OGLE LMC-ECL-15664}     &              &80.7443.1746& 05:26:21.14 & -68:46:44.5 & 15.91 &  0.07 &-0.273 &   0.024& -0.313   \\ 
 \object{OGLE LMC-ECL-15764}     &   SC4 281015 &77.7547.1048& 05:26:35.74 & -69:55:54.6 & 18.03 &   --  &    -- &   0.030& -0.059   \\ 
 \object{OGLE LMC-ECL-15779}     &   SC4 296152 &77.7548.325 & 05:26:36.54 & -69:51:25.4 & 16.74 &   --  &    -- &  -0.395&  0.099   \\ 
 \object{OGLE LMC-ECL-16178}     &              &            & 05:27:23.88 & -69:05:00.7 & 16.53 &  0.09 &-0.268 &  -0.085& -0.151   \\ 
 \object{OGLE LMC-ECL-16310}     &              &            & 05:27:41.15 & -69:16:20.3 & 15.57 & -0.03 &-0.298 &  -0.068& -0.122   \\ 
 \object{OGLE LMC-ECL-16320$^*$} &              & 7.7658.56  & 05:27:42.23 & -70:35:59.6 & 15.93 &   --  &    -- &  -0.086& -0.348   \\ 
 \object{OGLE LMC-ECL-16414}     &              &            & 05:27:53.17 & -69:03:51.6 & 16.53 &  0.16 &-0.189 &  -0.216& -0.073   \\ 
 \object{OGLE LMC-ECL-16442}     &    SC3 46420 &77.7669.3224& 05:27:56.24 & -69:53:12.3 & 19.03 &   --  &    -- &   0.127& -0.155   \\ 
 \object{OGLE LMC-ECL-16625}     &              &            & 05:28:18.80 & -69:03:55.0 & 16.41 & -0.15 &-0.249 &   0.096& -0.178   \\ 
 \object{OGLE LMC-ECL-16732}     &              &            & 05:28:31.27 & -69:13:40.7 & 16.10 &  0.13 &-0.234 &   0.120& -0.260   \\ 
 \noalign{\smallskip}\hline
\end{tabular}}\\
\begin{minipage}{0.85\textwidth}
\scriptsize Note: The full name from the OGLE II survey should be OGLE LMC-SCn nnnnnn. [A]
Out-of-eclipse $V$ magnitude based on OGLE database, see \cite{2011AcA....61..103G} and
\cite{2016AcA....66..421P}; [B]  photometric index by \cite{2002ApJS..141...81M}; [C]  photometric
index by \cite{2002AJ....123..855Z}. An asterisk in the name of particular system flags that this
system has dubious temperature $T_1$ as derived from its photometric index $(B-V)_0$. Therefore,
for these stars their effective temperatures were only roughly estimated, see the text for details.
\end{minipage}
\end{table*}

\begin{table*}
\caption{Relevant information for the analysed systems - continuation.} \label{InfoSystems2}
 \scriptsize
  \centering \scalebox{0.99}{
\begin{tabular}{lccccccccccc}
   \hline\hline\noalign{\smallskip}
 \multicolumn{3}{c}{S\,y\,s\,t\,e\,m\,\,\,\, n\,a\,m\,e} &   RA   &   DE     & $V_{\rm max}^{\,A}$ &  $(B-V)^{B}$ & $(B-V)_0^{B}$ & $(B-V)^{C}$ & $(B-V)_0^{C}$ \\
 \multicolumn{1}{c}{OGLE III}& OGLE II     &  MACHO      &        &          & [mag]               &    [mag]     &   [mag]       &   [mag]     & [mag]         \\
    \hline\noalign{\smallskip}
 \object{OGLE LMC-ECL-16925}     &             &             & 05:28:52.91 & -70:20:52.6 & 17.05 &   --  &   --  &  0.036 & -0.256 \\ 
 \object{OGLE LMC-ECL-16928}     &  SC3 317490 &             & 05:28:53.03 & -69:32:27.2 & 18.09 &   --  &   --  &  0.062 & -0.056 \\ 
 \object{OGLE LMC-ECL-16983}     &  SC3 329044 & 77.7917.602 & 05:29:00.17 & -69:25:53.1 & 18.17 &   --  &   --  & -0.005 & -0.128 \\ 
 \object{OGLE LMC-ECL-17030}     &  SC3 317678 & 77.7916.397 & 05:29:06.00 & -69:31:16.4 & 17.59 &   --  &   --  &  0.040 & -0.137 \\ 
 \object{OGLE LMC-ECL-17042}     &             & 82.7926.48  & 05:29:07.25 & -68:53:02.3 & 16.47 & -0.03 &-0.212 & -0.192 & -0.005 \\ 
 \object{OGLE LMC-ECL-17183}     &             & 7.7903.453  & 05:29:23.15 & -70:22:28.0 & 17.80 &   --  &    -- & -0.004 & -0.055 \\ 
 \object{OGLE LMC-ECL-17198}     &             &             & 05:29:24.63 & -68:42:13.1 & 14.12 & -0.17 &-0.268 & -0.048 & -0.241 \\ 
 \object{OGLE LMC-ECL-17223}     &             &             & 05:29:28.97 & -68:44:10.8 & 16.54 & -0.26 &-0.184 & -0.114 & -0.213 \\ 
 \object{OGLE LMC-ECL-17226$^*$} &             &             & 05:29:29.33 & -71:01:18.7 & 18.80 &   --  &    -- &  0.612 & -0.522 \\ 
 \object{OGLE LMC-ECL-17236}     &             &             & 05:29:30.94 & -68:40:12.8 & 17.75 &   --  &    -- & -0.194 & -0.042 \\ 
 \object{OGLE LMC-ECL-17267}     &             & 82.8049.37  & 05:29:34.38 & -68:42:33.3 & 15.89 & -0.13 &-0.257 & -0.124 & -0.283 \\ 
 \object{OGLE LMC-ECL-17283}     &             & 21.8008.68  & 05:29:37.25 & -71:26:43.4 & 17.34 &   --  &    -- & -0.011 & -0.151 \\ 
 \object{OGLE LMC-ECL-17299}     &  SC3 402931 & 77.8034.941 & 05:29:39.38 & -69:44:46.3 & 18.11 &   --  &    -- &  0.006 & -0.039 \\ 
 \object{OGLE LMC-ECL-17406}     &             &             & 05:29:51.43 & -68:39:53.3 & 16.07 & -0.17 &-0.218 & -0.084 & -0.260 \\ 
 \object{OGLE LMC-ECL-17498}     &             & 7.8025.972  & 05:30:01.77 & -70:21:22.1 & 17.91 &   --  &    -- & -0.059 & -0.167 \\ 
 \object{OGLE LMC-ECL-17543}     &             & 82.8047.152 & 05:30:07.32 & -68:50:34.4 & 17.22 &  0.16 &-0.115 & -0.027 & -0.149 \\ 
 \object{OGLE LMC-ECL-17579}     &             &             & 05:30:12.26 & -68:30:32.6 & 15.78 &  0.03 &-0.246 &  0.013 & -0.235 \\ 
 \object{OGLE LMC-ECL-17711}     &  SC2 103719 & 77.8158.89  & 05:30:28.68 & -69:30:03.7 & 16.57 & -0.17 &-0.212 & -0.027 & -0.262 \\ 
 \object{OGLE LMC-ECL-17718}     &             & 7.8140.136  & 05:30:29.24 & -70:43:32.1 & 17.11 &   --  &    -- & -0.084 & -0.083 \\ 
 \object{OGLE LMC-ECL-17774}     &             &             & 05:30:35.11 & -69:03:33.3 & 14.91 & -0.06 &-0.271 & -0.100 & -0.234 \\ 
 \object{OGLE LMC-ECL-17777}     &             & 82.8170.32  & 05:30:35.46 & -68:42:45.5 & 16.36 &  0.01 &-0.215 & -0.026 & -0.234 \\ 
 \object{OGLE LMC-ECL-17800}     &  SC2 214143 & 82.8160.2708& 05:30:39.35 & -69:23:57.1 & 16.97 &   --  &    -- & -0.037 & -0.186 \\ 
 \object{OGLE LMC-ECL-17809}     &             & 21.8126.151 & 05:30:40.46 & -71:40:32.0 & 17.72 &   --  &    -- & -0.032 & -0.048 \\ 
 \object{OGLE LMC-ECL-17923}     &             &  4.8177.63  & 05:30:54.30 & -68:16:17.4 & 17.48 &  0.16 &-0.155 &  1.371 & -0.494 \\ 
 \object{OGLE LMC-ECL-18077}     &             &             & 05:31:11.71 & -71:04:21.2 & 18.02 &   --  &    -- &  0.060 & -0.085 \\ 
 \object{OGLE LMC-ECL-18152}     &             &             & 05:31:21.80 & -69:11:15.6 & 16.15 &  0.16 &-0.275 &  0.229 & -0.306 \\ 
 \object{OGLE LMC-ECL-18174}     &  SC2 315183 & 77.8281.58  & 05:31:24.61 & -69:25:28.1 & 16.96 & -0.13 &-0.075 &  0.051 & -0.165 \\ 
 \object{OGLE LMC-ECL-18316$^*$} &             &             & 05:31:43.49 & -68:28:10.5 & 17.20 &   --  &    -- &  0.408 & -0.419 \\ 
 \object{OGLE LMC-ECL-18355$^*$} &             &             & 05:31:47.85 & -68:33:26.7 & 16.24 &   --  &    -- & -0.938 &  0.318 \\ 
 \object{OGLE LMC-ECL-18371}     &             & 8.8424.131  & 05:31:49.93 & -67:53:52.1 & 18.05 &   --  &    -- & -0.002 & -0.133 \\ 
 \object{OGLE LMC-ECL-18408}     &             &             & 05:31:54.55 & -68:59:19.7 & 17.38 &  0.42 &-0.061 &  0.578 & -0.383 \\ 
 \object{OGLE LMC-ECL-18500}     &  SC2 351438 &             & 05:32:06.62 & -70:04:56.5 & 17.97 &   --  &    -- &  0.235 & -0.137 \\ 
 \object{OGLE LMC-ECL-18501$^*$} &             & 82.8407.1180& 05:32:06.68 & -69:04:34.5 & 18.00 &   --  &    -- &  0.374 & -0.337 \\ 
 \object{OGLE LMC-ECL-18579}     &             & 82.8404.209 & 05:32:16.49 & -69:15:35.7 & 17.03 &  0.17 &-0.138 &  0.010 & -0.216 \\ 
 \object{OGLE LMC-ECL-19095}     &             &             & 05:33:26.76 & -69:10:51.3 & 16.63 & -0.03 &-0.248 & -0.069 & -0.137 \\ 
 \object{OGLE LMC-ECL-19294}     &             & 82.8648.78  & 05:33:53.71 & -69:09:02.8 & 16.89 &  0.54 &-0.189 & -0.023 & -0.201 \\ 
 \object{OGLE LMC-ECL-19577}     &  SC1 335438 & 81.8760.105 & 05:34:32.08 & -69:45:36.2 & 16.08 &   --  &    -- & -0.081 & -0.218 \\ 
 \object{OGLE LMC-ECL-19583}     &             & 14.8744.3528& 05:34:33.06 & -70:49:10.4 & 17.78 &   --  &    -- &  0.046 & -0.240 \\ 
 \object{OGLE LMC-ECL-19624}     &             & 8.8780.16   & 05:34:39.71 & -68:21:55.5 & 15.94 & -0.19 &-0.280 & -0.433 & -0.153 \\ 
 \object{OGLE LMC-ECL-19675}     &             &             & 05:34:48.91 & -68:19:46.7 & 17.13 & -0.26 &-0.105 & -0.124 & -0.177 \\ 
 \object{OGLE LMC-ECL-19727$^*$} &             &             & 05:34:57.01 & -69:15:41.6 & 18.11 &   --  &    -- &  0.434 & -0.455 \\ 
 \object{OGLE LMC-ECL-19879}     &  SC16 70662 & 81.8881.47  & 05:35:17.59 & -69:43:18.8 & 15.02 &   --  &    -- &  0.014 & -0.296 \\ 
 \object{OGLE LMC-ECL-20053}     &             &             & 05:35:41.09 & -68:58:57.9 & 16.10 &  0.07 &-0.213 &  0.117 & -0.237 \\ 
 \object{OGLE LMC-ECL-20066}     &             &             & 05:35:42.99 & -69:14:10.4 & 17.24 &   --  &    -- &  0.004 & -0.176 \\ 
 \object{OGLE LMC-ECL-20285}     &             &             & 05:36:09.54 & -68:55:27.4 & 16.29 &   --  &    -- & -0.480 & -0.012 \\ 
 \object{OGLE LMC-ECL-20299}     &             & 81.9005.17  & 05:36:10.74 & -69:32:02.2 & 14.63 & -0.08 &-0.306 & -0.019 & -0.299 \\ 
 \object{OGLE LMC-ECL-20384}     &             & 82.9132.83  & 05:36:20.35 & -69:06:15.7 & 16.57 &  0.10 &-0.283 & -0.061 & -0.178 \\ 
 \object{OGLE LMC-ECL-20550}     &             & 82.9132.145 & 05:36:43.34 & -69:06:07.1 & 17.03 & -0.34 &-0.116 & -0.352 & -0.090 \\ 
 \object{OGLE LMC-ECL-20589}     &             & 82.9130.25  & 05:36:48.74 & -69:16:59.4 & 15.19 & -0.13 &-0.271 & -0.109 & -0.292 \\ 
 \object{OGLE LMC-ECL-20648}     &             & 81.9127.131 & 05:36:56.90 & -69:28:50.3 & 16.65 &  0.09 &-0.185 & -0.010 & -0.235 \\ 
 \object{OGLE LMC-ECL-20735}     & SC16 240980 & 81.9238.590 & 05:37:07.73 & -70:09:38.3 & 17.88 &   --  &    -- &  0.092 &  0.021 \\ 
 \object{OGLE LMC-ECL-20783$^*$} & SC16 226238 & 11.9235.33  & 05:37:14.02 & -70:20:01.5 & 15.67 &   --  &    -- &  0.329 & -0.409 \\ 
 \object{OGLE LMC-ECL-20860}     &             & 81.9246.29  & 05:37:24.44 & -69:34:19.6 & 15.87 &   --  &    -- & -0.222 & -0.173 \\ 
 \object{OGLE LMC-ECL-21259}     &             &             & 05:38:20.42 & -69:03:10.4 & 15.98 &  0.69 &-0.430 &  0.296 & -0.267 \\ 
 \object{OGLE LMC-ECL-21278}     &             &             & 05:38:22.59 & -69:21:04.5 & 17.38 &   --  &    -- &  0.361 & -0.272 \\ 
 \object{OGLE LMC-ECL-21479}     &             &             & 05:38:46.72 & -69:02:40.5 & 16.84 &   --  &    -- & -0.451 & -0.038 \\ 
 \object{OGLE LMC-ECL-21647}     &             & 50.9634.17  & 05:39:12.03 & -67:55:54.1 & 16.54 & -0.08 &-0.173 & -0.014 & -0.200 \\ 
 \object{OGLE LMC-ECL-21695}     &             &             & 05:39:22.14 & -68:31:58.2 & 16.85 & -0.09 &-0.181 & -0.304 & -0.142 \\ 
 \object{OGLE LMC-ECL-21961$^*$} &             &             & 05:40:02.01 & -69:22:10.6 & 17.48 &   --  &    -- &  0.548 & -0.445 \\ 
 \object{OGLE LMC-ECL-21968}     &             &             & 05:40:03.22 & -69:19:54.7 & 15.90 &   --  &    -- &  0.041 & -0.244 \\ 
 \object{OGLE LMC-ECL-22069}     &             &             & 05:40:17.90 & -69:31:31.8 & 17.65 &   --  &    -- &  0.393 & -0.308 \\ 
 \object{OGLE LMC-ECL-22211$^*$} &             &76.9730.1611 & 05:40:38.42 & -69:36:22.6 & 17.37 &   --  &    -- &  0.224 & -0.345 \\ 
 \object{OGLE LMC-ECL-22232}     & SC18 107244 & 76.9845.63  & 05:40:41.43 & -69:59:01.4 & 16.39 &   --  &    -- & -0.040 & -0.212 \\ 
 \object{OGLE LMC-ECL-22422$^*$} &             &             & 05:41:10.20 & -69:19:30.3 & 17.97 &   --  &    -- &  0.284 & -0.316 \\ 
 \object{OGLE LMC-ECL-22455}     &  SC18 66331 & 11.9835.112 & 05:41:13.80 & -70:39:49.5 & 17.18 &   --  &    -- &  0.105 & -0.210 \\ 
 \object{OGLE LMC-ECL-22494}     &             & 76.9849.541 & 05:41:19.17 & -69:41:43.3 & 18.59 &   --  &    -- &  0.088 & -0.129 \\ 
 \object{OGLE LMC-ECL-22613$^*$} &             &  15.9946.7  & 05:41:37.74 & -71:19:02.8 & 14.77 &   --  &    -- &  0.304 & -0.529 \\ 
 \object{OGLE LMC-ECL-22695$^*$} &             &  76.9970.41 & 05:41:50.30 & -69:42:24.5 & 16.13 &   --  &    -- & -1.182 &  0.522 \\ 
 \object{OGLE LMC-ECL-23298}     &             &             & 05:43:26.23 & -69:16:25.5 & 15.47 &  0.01 &-0.275 &  0.118 & -0.279 \\ 
 \object{OGLE LMC-ECL-23323}     &             &             & 05:43:32.02 & -69:13:40.6 & 17.49 &   --  &    -- & -0.007 & -0.155 \\ 
 \object{OGLE LMC-ECL-23920}     &             &             & 05:45:54.86 & -69:13:44.1 & 17.53 &   --  &    -- &  0.070 & -0.189 \\ 
 \object{OGLE LMC-ECL-24112}     &             & 76.10813.52 & 05:46:44.34 & -70:00:37.8 & 16.69 &   --  &    -- &  0.085 & -0.153 \\ 
 \object{OGLE LMC-ECL-24236$^*$} &             &             & 05:47:18.17 & -69:39:18.5 & 17.55 &   --  &    -- &  0.168 & -0.333 \\ 
 \object{OGLE LMC-ECL-24534}     &             &             & 05:48:43.17 & -70:01:00.3 & 16.24 &   --  &    -- & -0.033 & -0.292 \\ 
 \object{OGLE LMC-ECL-24817}     &             &             & 05:50:12.61 & -69:55:26.4 & 18.10 &   --  &    -- & -0.287 & -0.085 \\ 
 \object{OGLE LMC-ECL-25047}     &             &             & 05:51:51.36 & -70:01:21.1 & 18.05 &   --  &    -- & -0.072 & -0.213 \\ 
 \object{OGLE LMC-ECL-25227}     &             & 30.11783.137& 05:53:17.33 & -69:53:13.2 & 18.02 &   --  &    -- & -0.067 & -0.183 \\ 
 \object{OGLE LMC-ECL-25743$^*$} &             &             & 06:04:25.47 & -68:55:28.1 & 15.29 &   --  &    -- &  0.105 & -0.417 \\ 
 \object{OGLE LMC-ECL-25885}     &             &             & 06:07:52.60 & -71:46:42.9 & 17.88 &   --  &    -- &  0.059 & -0.023 \\ 
 \object{OGLE LMC-ECL-25980$^*$} &             &             & 06:10:48.42 & -69:52:17.5 & 17.09 &   --  &    -- &    --  &     -- \\ 
 \noalign{\smallskip}\hline
\end{tabular}}\\
\end{table*}

\begin{table*}
 \caption{Parameters of the light curve fits and the apsidal motion.}
 \label{LCOCparam}
 \scriptsize
 \centering \scalebox{0.78}{
 \begin{tabular}{l c r c c c c c c c r c c c c c c c}
 \hline\hline
  System            &    $i$       &  \mlc{$T_1$}  & \mlc{$T_2$} &   $L_1$   &   $L_2$   &   $L_3$   &  $R_1/a$  &  $R_2/a$  & $HJD_0     $   &  \mlc{$P$ [d]} &   $e$     & $\omega$ [deg] & $U$ [yr] \\
                    &    [deg]     &  \mlc{[K]}    & \mlc{ [K]}  &   [\%]    &    [\%]   &   [\% ]   &           &           & $[2450000+]$   &                &           &                &          \\
 \hline
 OGLE LMC-ECL-00127 & 80.31 (0.54) & 12500 (fixed) & 12153 (189) & 57.6 (1.9)& 42.3 (1.7)& 0         & 0.252 (5) & 0.230 (3) & 3400.7935 (24) & 1.9716646 (19) & 0.020 (8) & 118.3 (1.2) & 47.8 (4.9)  \\ 
 OGLE LMC-ECL-00442 & 87.02 (1.12) & 29000 (fixed) & 18691 (302) & 78.7 (2.3)& 11.3 (1.2)& 10.0 (0.7)& 0.194 (3) & 0.108 (4) & 5552.2497 (115)& 3.4108511 (123)& 0.341 (17)&  49.7 (7.8) & 94.1 (10.3) \\ 
 OGLE LMC-ECL-00483 & 86.94 (0.96) & 15700 (fixed) &  9767 (619) & 70.4 (7.0)& 12.3 (2.3)& 17.3 (5.2)& 0.197 (11)& 0.118 (9) & 5551.8390 (89) & 1.9408981 (57) & 0.139 (26)& 186.9 (15.3)& 77.0 (24.7) \\ 
 OGLE LMC-ECL-00510 & 83.71 (0.67) &  8270 (fixed) &  8179 (118) & 50.7 (2.1)& 46.3 (0.8)& 2.9 (1.3) & 0.228 (3) & 0.217 (2) & 5552.2803 (92) & 2.4828516 (130)& 0.096 (12)& 171.9 (6.7) &106.7 (18.9) \\ 
 OGLE LMC-ECL-00527 & 80.68 (0.35) &  8150 (fixed) &  7388 (103) & 63.2 (1.5)& 36.7 (1.4)& 0         & 0.227 (2) & 0.194 (2) & 5552.5222 (21) & 1.7258405 (14) & 0.045 (16)& 106.0 (2.3) & 57.9 (11.5) \\ 
 OGLE LMC-ECL-00648 & 84.13 (0.64) & 16200 (fixed) & 14675 (257) & 77.3 (2.6)& 20.1 (1.3)& 2.7 (2.0) & 0.276 (3) & 0.151 (4) & 5553.3481 (33) & 1.5805259 (29) & 0.046 (11)& 106.1 (2.0) & 26.8 (9.0)  \\ 
 OGLE LMC-ECL-00652 & 86.34 (0.29) & 20600 (fixed) & 21103 (222) & 25.1 (1.2)& 74.9 (2.0)& 0         & 0.139 (2) & 0.232 (3) & 5553.0384 (102)& 3.2610612 (213)& 0.202 (21)&  68.1 (9.4) & 92.1 (12.6) \\ 
 OGLE LMC-ECL-00666 & 88.21 (0.56) & 30000 (fixed) & 21888 (511) & 45.4 (3.2)& 12.7 (1.3)&41.9 (2.8) & 0.247 (5) & 0.163 (4) & 5552.9386 (24) & 1.6029021 (20) & 0.036 (10)& 162.6 (1.8) & 39.9 (5.2)  \\ 
 OGLE LMC-ECL-00737 & 89.27 (0.24) & 20600 (fixed) & 20615 (210) & 49.0 (0.4)& 51.0 (0.5)& 0         & 0.209 (3) & 0.211 (3) & 5601.1048 (45) & 1.9848417 (86) & 0.111 (32)& 310.5 (3.4) & 44.1 (8.9)  \\ 
 OGLE LMC-ECL-00868 & 85.18 (0.71) & 10700 (fixed) & 11476 (234) & 35.3 (1.8)& 59.6 (2.1)& 5.1 (1.2) & 0.148 (2) & 0.183 (3) & 5600.2204 (123)& 2.2747035 (224)& 0.203 (36)& 167.1 (12.9)&114.8 (18.2) \\ 
 OGLE LMC-ECL-00872 & 88.29 (0.85) & 17800 (fixed) & 14181 (427) & 67.2 (0.3)& 32.8 (0.3)& 0         & 0.227 (3) & 0.186 (4) & 5600.0326 (37) & 1.2823707 (23) & 0.071 (17)& 127.6 (2.5) & 32.8 (1.8)  \\ 
 OGLE LMC-ECL-00929 & 89.53 (0.59) &  8170 (fixed) &  7890 (208) & 56.0 (2.5)& 44.0 (2.2)& 0         & 0.177 (5) & 0.163 (9) & 5600.4775 (78) & 1.8515324 (103)& 0.210 (45)& 262.5 (12.0)&103.9 (19.3) \\ 
 OGLE LMC-ECL-00955 & 85.37 (0.37) & 14000 (fixed) & 17175 (536) & 35.6 (1.7)& 60.3 (0.7)& 4.1 (2.9) & 0.144 (5) & 0.162 (6) & 5599.7752 (146)& 2.8025063 (221)& 0.209 (59)& 126.6 (11.9)&195.1 (33.0) \\ 
 OGLE LMC-ECL-01445 & 86.61 (0.22) & 12500 (fixed) & 12380 (155) & 57.8 (1.2)& 32.6 (0.5)& 9.6 (6.1) & 0.236 (2) & 0.179 (2) & 6001.9480 (89) & 2.2912938 (68) & 0.129 (20)& 164.4 (7.1) & 54.5 (5.8)  \\ 
 OGLE LMC-ECL-02912 & 78.18 (0.52) & 14000 (fixed) & 13031 (213) & 58.4 (2.0)& 39.0 (1.8)& 2.6 (1.8) & 0.224 (4) & 0.195 (3) & 6001.7281 (14) & 2.1106452 (17) & 0.020 (11)& 264.0 (17.2)& 71.9 (18.3) \\ 
 OGLE LMC-ECL-02943 & 81.87 (0.47) & 10700 (fixed) & 10044 (135) & 60.2 (1.3)& 38.8 (1.7)& 0.9 (0.8) & 0.203 (7) & 0.167 (5) & 6002.3439 (33) & 3.2126214 (30) & 0.090 (23)& 100.1 (1.8) &112.8 (13.2) \\ 
 OGLE LMC-ECL-04957 & 89.33 (0.43) & 10000 (fixed) &  9383 (115) & 60.8 (1.4)& 39.2 (1.0)& 0         & 0.224 (2) & 0.189 (2) & 5003.0905 (23) & 1.4541166 (11) & 0.065 (8) &  87.0 (2.4) & 37.4 (3.9)  \\ 
 OGLE LMC-ECL-05345 & 83.85 (0.27) & 26000 (fixed) & 26050 (288) & 56.0 (0.9)& 43.4 (1.1)& 0.6 (0.5) & 0.182 (2) & 0.162 (3) & 5002.5639 (91) & 4.0348631 (121)& 0.261 (22)& 137.2 (6.0) &131.4 (14.7) \\ 
 OGLE LMC-ECL-06837 & 83.10 (0.39) & 22500 (fixed) & 21103 (319) & 54.8 (1.5)& 45.2 (1.2)& 0         & 0.175 (4) & 0.165 (3) & 6002.6578 (52) & 2.7470370 (42) & 0.048 (7) & 104.8 (1.3) &  9.9 (0.6)  \\ 
 OGLE LMC-ECL-07578 & 88.03 (0.48) & 20600 (fixed) & 19882 (257) & 52.2 (0.9)& 28.7 (1.0)& 19.1 (2.3)& 0.211 (4) & 0.186 (7) & 5002.8096 (78) & 4.0185993 (98) & 0.141`(15)& 316.8 (7.9) &146.0 (15.2) \\ 
 OGLE LMC-ECL-07585 & 88.88 (0.54) & 11000 (fixed) &  9850 (210) & 54.3 (1.6)& 45.6 (1.4)& 0         & 0.201 (5) & 0.198 (4) & 5000.7113 (34) & 1.5331843 (40) & 0.059 (22)& 302.7 (18.8)& 39.1 (22.3) \\ 
 OGLE LMC-ECL-07641 & 82.32 (0.80) & 18500 (fixed) & 19364 (776) & 33.7 (1.2)& 50.2 (2.2)& 16.1 (3.4)& 0.154 (7) & 0.181 (6) & 5495.0273 (112)& 6.3299708 (230)& 0.265 (110)& 89.8 (9.0) &198.9 (52.6) \\ 
 OGLE LMC-ECL-07838 & 87.69 (1.09) & 17000 (fixed) & 16180 (328) & 63.6 (3.0)& 35.2 (1.6)& 1.2 (1.2) & 0.233 (4) & 0.169 (5) & 4002.2882 (61) & 2.6211802 (81) & 0.123 (43)& 102.2 (2.7) & 83.3 (10.9) \\ 
 OGLE LMC-ECL-08311 & 75.76 (0.31) & 23500 (fixed) & 17805 (241) & 64.4 (1.1)& 35.6 (0.7)& 0         & 0.219 (2) & 0.202 (3) & 4001.6423 (53) & 2.3454109 (42) & 0.107 (12)&  84.7 (4.4) & 35.4 (1.3)  \\ 
 OGLE LMC-ECL-08397 & 83.80 (0.62) & 30000 (fixed) & 26080 (136) & 42.8 (5.2)& 30.7 (3.1)& 26.5 (4.6)& 0.207 (7) & 0.200 (4) & 4003.0302 (107)& 3.1317961 (170)& 0.169 (21)& 138.9 (7.1) & 58.6 (3.4)  \\ 
 OGLE LMC-ECL-08439 & 86.11 (1.08) & 16700 (fixed) & 16657 (505) & 42.7 (3.3)& 37.7 (2.0)& 19.6 (4.8)& 0.247 (6) & 0.233 (3) & 4002.1008 (32) & 2.0488466 (12) & 0.040 (7) &  71.0 (1.6) & 39.5 (3.1)  \\ 
 OGLE LMC-ECL-08752 & 80.74 (0.35) & 31500 (fixed) & 24970 (213) & 62.7 (0.9)& 37.3 (0.7)& 0         & 0.221 (3) & 0.211 (3) & 6157.0514 (44) & 2.6788383 (65) & 0.039 (9) &  14.1 (2.2) & 49.6 (5.0)  \\ 
 OGLE LMC-ECL-08859 & 77.79 (0.59) & 10000 (fixed) &  8721 (234) & 51.8 (1.4)& 38.1 (1.6)& 10.1 (3.5)& 0.245 (6) & 0.233 (5) & 4001.5476 (9)  & 1.0877549 (5)  & 0.046 (4) & 249.4 (1.1) & 15.7 (1.2)  \\ 
 OGLE LMC-ECL-09186 & 81.78 (0.12) & 24500 (fixed) & 28734 (108) & 41.6 (0.8)& 58.4 (0.9)& 0         & 0.182 (3) & 0.186 (4) & 5002.6794 (69) & 3.7734723 (141)& 0.236 (16)& 148.1 (5.9) &126.1 (34.9) \\ 
 OGLE LMC-ECL-10302 & 75.99 (1.27) & 24500 (fixed) & 22720 (297) & 25.0 (0.3)& 19.8 (0.4)& 55.2 (1.4)& 0.234 (2) & 0.220 (3) & 4001.0612 (6)  & 1.5401250 (12) & 0.008 (2) &  35.2 (6.0) & 17.9 (1.4)  \\ 
 OGLE LMC-ECL-10377 & 85.42 (0.46) & 20600 (fixed) & 17949 (132) & 56.7 (2.7)& 43.3 (2.5)& 0         & 0.240 (9) & 0.214 (8) & 2246.1620 (8)  & 2.1087293 (11) & 0.030 (4) & 329.8 (4.3) & 32.7 (2.1)  \\ 
 OGLE LMC-ECL-10446 & 80.17 (0.91) & 10000 (fixed) &  9025 (204) & 59.3 (1.6)& 36.4 (2.1)& 4.3 (2.7) & 0.230 (11)& 0.193 (9) & 5001.3796 (59) & 2.1452228 (77) & 0.096 (13)& 146.9 (4.8) & 36.4 (8.5)  \\ 
 OGLE LMC-ECL-10601 & 79.62 (0.28) & 17000 (fixed) & 14687 (190) & 65.7 (0.6)& 34.3 (0.5)& 0         & 0.205 (3) & 0.164 (3) & 4002.0628 (179)& 3.2837060 (242)& 0.225 (63)& 156.0 (11.7)&119.4 (28.9) \\ 
 OGLE LMC-ECL-10672 & 88.34 (0.83) & 20600 (fixed) & 17942 (275) & 65.7 (2.4)& 32.6 (1.9)& 1.7 (1.4) & 0.240 (4) & 0.187 (5) & 4002.4626 (11) & 1.7056091 (14) & 0.031 (9) & 293.3 (12.9)& 52.8 (19.3) \\ 
 OGLE LMC-ECL-10867 & 82.86 (0.25) & 18500 (fixed) & 17569 (174) & 63.7 (2.1)& 36.3 (1.7)& 0         & 0.236 (5) & 0.176 (7) & 4002.4395 (19) & 1.8582461 (25) & 0.043 (11)& 296.4 (7.2) & 60.5 (11.0) \\ 
 OGLE LMC-ECL-11183 & 87.77 (1.07) & 16000 (fixed) & 26957 (503) &  5.7 (0.8)& 48.4 (8.2)& 45.9 (7.0)& 0.155 (8) & 0.298 (11)& 4000.9668 (85) & 2.5808061 (107)& 0.199 (13)&  88.5 (9.4) & 48.1 (7.6)  \\ 
 OGLE LMC-ECL-11320 & 84.51 (0.30) & 14500 (fixed) & 14410 (162) & 63.3 (1.5)& 34.1 (1.3)& 2.6 (2.1) & 0.243 (2) & 0.178 (2) & 4002.9282 (30) & 2.5380708 (62) & 0.052 (16)& 123.0 (11.8)& 74.5 (15.2) \\ 
 OGLE LMC-ECL-11351 & 77.87 (0.29) & 17000 (fixed) & 20072 (266) & 52.0 (0.7)& 48.0 (0.6)& 0         & 0.212 (3) & 0.181 (3) & 4002.9247 (72) & 2.4103248 (69) & 0.150 (23)&  64.7 (5.0) & 92.5 (13.7) \\ 
 OGLE LMC-ECL-11374 & 87.94 (0.32) & 12500 (fixed) & 12544 (137) & 55.3 (1.4)& 43.8 (1.7)& 0.9 (0.8) & 0.231 (4) & 0.210 (5) & 4003.4547 (25) & 2.0154752 (26) & 0.045 (12)&  23.9 (3.1) & 88.0 (22.8) \\ 
 OGLE LMC-ECL-11680 & 81.45 (0.35) & 10000 (fixed) & 10386 (177) & 48.6 (0.9)& 51.4 (1.1)& 0         & 0.183 (2) & 0.185 (2) & 4001.6785 (56) & 2.1649015 (68) & 0.131 (30)&  61.1 (4.0) & 72.0 (5.1)  \\ 
 OGLE LMC-ECL-11854 & 80.37 (0.19) & 17000 (fixed) & 16556 (252) & 72.9 (2.6)& 24.4 (0.9)& 2.7 (1.7) & 0.259 (3) & 0.153 (3) & 5001.3761 (18) & 2.4000588 (35) & 0.041 (12)&  71.5 (2.9) & 66.3 (4.3)  \\ 
 OGLE LMC-ECL-11929 & 75.30 (0.48) & 23000 (fixed) & 21808 (119) & 66.1 (1.7)& 33.1 (0.8)& 0.8 (0.8) & 0.281 (5) & 0.210 (4) & 5000.9802 (61) & 4.4228109 (101)& 0.050 (13)& 355.5 (4.5) & 87.8 (5.0)  \\ 
 OGLE LMC-ECL-12002 & 81.65 (0.17) & 20600 (fixed) & 18815 (185) & 57.7 (0.7)& 42.3 (0.8)& 0         & 0.208 (4) & 0.192 (5) & 4000.5253 (40) & 2.1911826 (129)& 0.064 (19)& 141.6 (3.8) & 57.4 (9.2)  \\ 
 OGLE LMC-ECL-12043 & 81.71 (2.30) & 14500 (fixed) & 15377 (395) & 31.5 (2.2)& 41.2 (3.0)& 27.3 (4.9)& 0.199 (12)& 0.205 (11)& 4002.7978 (41) & 1.3432745 (73) & 0.150 (14)& 146.1 (2.3) & 48.7 (4.6)  \\ 
 OGLE LMC-ECL-12234 & 83.48 (0.38) & 16000 (fixed) & 15342 (240) & 68.7 (0.6)& 29.2 (0.7)& 2.1 (1.4) & 0.252 (2) & 0.171 (2) & 5003.9357 (68) & 2.5755734 (91) & 0.108 (15)& 205.7 (3.0) & 87.9 (2.1)  \\ 
 OGLE LMC-ECL-12323 & 86.39 (0.21) & 20600 (fixed) & 20872 (351) & 51.3 (1.3)& 45.8 (2.2)& 2.9 (2.5) & 0.146 (2) & 0.142 (3) & 4002.3660 (162)& 2.9617909 (164)& 0.319 (22)&  61.9 (12.7)&183.7 (33.9) \\ 
 OGLE LMC-ECL-12504 & 80.94 (0.33) & 15000 (fixed) & 18530 (313) & 23.7 (3.1)& 44.3 (3.7)& 32.0 (5.0)& 0.181 (4) & 0.217 (5) & 4002.5246 (59) & 2.3873668 (117)& 0.175 (41)&  65.0 (4.8) &159.4 (48.0) \\ 
 OGLE LMC-ECL-12513 & 82.04 (0.75) & 14000 (fixed) & 12527 (329) & 31.3 (1.5)& 17.0 (0.9)& 51.7 (6.2)& 0.212 (9) & 0.164 (6) & 3539.6785 (38) & 1.6462861 (32) & 0.110 (27)& 109.1 (3.2) & 44.1 (9.4)  \\ 
 OGLE LMC-ECL-12687 & 79.39 (0.37) & 10400 (fixed) &  9309 (188) & 60.8 (2.0)& 39.2 (1.8)& 0         & 0.244 (5) & 0.213 (3) & 4003.0971 (11) & 1.6872665 (10) & 0.031 (4) &  19.3 (4.6) & 32.2 (2.7)  \\ 
 OGLE LMC-ECL-12792 & 83.98 (0.82) &  8300 (fixed) &  8179 (214) & 50.7 (2.4)& 48.4 (1.1)& 0.9 (0.8) & 0.246 (7) & 0.238 (4) & 4002.9299 (14) & 1.7357179 (88) & 0.047 (21)& 253.9 (3.7) &194.2 (74.8) \\ 
 OGLE LMC-ECL-13039 & 84.12 (0.34) & 13500 (fixed) & 13557 (156) & 37.0 (0.5)& 63.0 (0.5)& 0         & 0.172 (2) & 0.223 (2) & 4003.1363 (29) & 1.5053762 (31) & 0.096 (5) &   9.1 (12.4)& 45.5 (3.1)  \\ 
 OGLE LMC-ECL-13198 & 78.55 (0.16) & 16000 (fixed) & 18256 (302) & 44.2 (0.6)& 54.9 (0.7)& 0.9 (0.5) & 0.201 (3) & 0.203 (4) & 4001.4109 (51) & 2.1174661 (74) & 0.114 (11)&  36.1 (4.9) & 63.0 (4.9)  \\ 
 OGLE LMC-ECL-13202 & 88.82 (0.32) & 10400 (fixed) & 11061 (135) & 42.6 (0.8)& 54.6 (1.1)& 2.8 (0.7) & 0.187 (2) & 0.202 (2) & 4002.3909 (67) & 2.2752552 (111)& 0.151 (32)&  66.0 (2.8) & 81.7 (11.0) \\ 
 OGLE LMC-ECL-13321 & 83.80 (0.19) & 18500 (fixed) & 21923 (238) & 41.8 (0.5)& 58.7 (0.9)& 0         & 0.185 (4) & 0.198 (9) & 4000.9603 (33) & 1.6517623 (42) & 0.160 (46)& 114.8 (2.1) & 63.0 (29.6) \\ 
 OGLE LMC-ECL-13407 & 80.31 (0.35) & 11500 (fixed) & 11130 (185) & 51.9 (0.7)& 48.1 (0.5)& 0         & 0.217 (4) & 0.215 (3) & 4001.9580 (26) & 1.9333573 (25) & 0.080 (6) &  96.7 (2.7) & 62.7 (5.2)  \\ 
 OGLE LMC-ECL-13452 & 85.36 (0.23) & 24500 (fixed) & 23186 (123) & 53.6 (1.0)& 46.3 (0.8)& 0         & 0.207 (3) & 0.201 (2) & 4002.0813 (37) & 1.9627478 (56) & 0.086 (4) & 333.4 (1.6) & 39.2 (2.5)  \\ 
 OGLE LMC-ECL-13467 & 85.71 (0.39) & 25200 (fixed) & 23906 (174) & 55.5 (1.6)& 43.7 (0.9)& 0.8 (0.6) & 0.231 (3) & 0.214 (3) & 4002.3241 (34) & 2.5059763 (61) & 0.063 (10)&  80.7 (3.0) & 46.9 (3.7)  \\ 
 OGLE LMC-ECL-13567 & 87.52 (0.48) &  8400 (fixed) &  7875  (49) & 59.4 (0.7)& 40.2 (0.6)& 0.5 (0.5) & 0.259 (2) & 0.227 (2) & 4001.6849 (9)  & 1.1749927 (8)  & 0.021 (2) &  83.2 (1.4) & 14.1 (1.1)  \\ 
 OGLE LMC-ECL-13601 & 80.49 (0.14) & 19000 (fixed) & 19481 (166) & 46.5 (0.6)& 55.5 (0.6)& 0         & 0.192 (2) & 0.202 (3) & 4003.4649 (25) & 2.5988993 (33) & 0.097 (22)& 100.1 (2.8) & 91.2 (12.3) \\ 
 OGLE LMC-ECL-13809 & 88.90 (0.19) & 18000 (fixed) & 16362 (145) & 47.6 (0.8)& 44.0 (1.0)& 8.2 (2.4) & 0.214 (3) & 0.213 (3) & 4002.1345 (27) & 1.7508997 (52) & 0.073 (25)&  68.7 (4.7) & 34.9 (10.8) \\ 
 OGLE LMC-ECL-13842 & 80.13 (0.25) & 26000 (fixed) & 19994 (117) & 62.8 (2.3)& 37.0 (1.4)& 0.2 (0.2) & 0.209 (4) & 0.201 (4) & 4002.1696 (66) & 2.2447821 (204)& 0.110 (33)& 128.9 (5.9) & 46.3 (13.0) \\ 
 OGLE LMC-ECL-13851 & 81.61 (0.16) & 17000 (fixed) & 15948 (102) & 53.8 (0.9)& 46.2 (0.7)& 0         & 0.230 (2) & 0.223 (3) & 4000.1045 (39) & 2.7544935 (64) & 0.080 (17)& 138.7 (3.6) & 69.5 (14.4) \\ 
 OGLE LMC-ECL-13902 & 81.11 (0.85) & 17000 (fixed) & 10077 (258) & 70.5 (2.8)& 19.6 (3.5)& 9.9 (0.7) & 0.238 (7) & 0.187 (9) & 4003.7461 (33) & 1.4033760 (69) & 0.107 (24)& 128.5 (2.5) & 31.6 (8.2)  \\ 
 OGLE LMC-ECL-14013 & 81.27 (0.72) & 26000 (fixed) & 21561 (847) & 46.6 (0.9)& 28.5 (0.8)& 24.9 (1.9)& 0.189 (3) & 0.176 (3) & 3000.9749 (130)& 2.2729766 (141)& 0.242 (77)&  35.9 (8.8) & 40.1 (11.1) \\ 
 OGLE LMC-ECL-14159 & 84.01 (0.38) & 16700 (fixed) & 14950 (159) & 58.3 (0.5)& 41.7 (0.4)& 0         & 0.179 (3) & 0.162 (4) & 5600.6748 (49) & 2.2433740 (50) & 0.125 (49)&  87.1 (4.0) &185.3 (77.9) \\ 
 OGLE LMC-ECL-14462 & 73.79 (0.89) & 20600 (fixed) & 16523 (320) & 54.6 (0.9)& 35.7 (0.7)& 9.7 (1.6) & 0.228 (7) & 0.216 (5) & 4002.8059 (24) & 1.8869262 (22) & 0.042 (11)& 185.7 (3.2) & 22.4 (2.0)  \\ 
 OGLE LMC-ECL-15161 & 78.11 (0.17) & 13200 (fixed) & 12197 (105) & 49.4 (0.5)& 43.3 (0.4)& 7.3 (0.7) & 0.235 (2) & 0.234 (2) & 4001.6262 (29) & 3.6255092 (73) & 0.090 (52)&  95.6 (13.8)&134.0 (34.7) \\ 
 OGLE LMC-ECL-15288 & 74.77 (0.22) & 20600 (fixed) & 20437 (123) & 43.3 (0.4)& 41.0 (0.5)& 15.7 (2.1)& 0.227 (3) & 0.222 (2) & 4001.2285 (31) & 2.6743442 (46) & 0.057 (10)&  79.9 (4.1) & 55.1 (2.8)  \\ 
 OGLE LMC-ECL-15386 & 81.25 (0.36) & 10400 (fixed) & 10300 (182) & 48.3 (0.7)& 47.8 (0.6)& 3.9 (1.2) & 0.228 (2) & 0.226 (2) & 4001.9444 (77) & 2.5117115 (116)& 0.120 (16)& 195.3 (8.4) & 84.1 (15.9) \\ 
 OGLE LMC-ECL-15441 & 83.94 (0.50) & 14000 (fixed) & 11324 (212) & 58.7 (1.1)& 38.5 (1.0)& 2.8 (1.4) & 0.167 (4) & 0.158 (3) & 4001.3043 (170)& 2.8857944 (168)& 0.280 (92)& 222.1 (12.7)&184.1 (29.6) \\ 
 OGLE LMC-ECL-15473 & 76.17 (0.43) & 12500 (fixed) & 13308 (270) & 37.2 (0.9)& 39.9 (1.8)& 22.9 (2.5)& 0.205 (2) & 0.206 (2) & 6001.2646 (41) & 1.5892106 (31) & 0.094 (11)& 111.5 (2.2) & 48.2 (3.0)  \\ 
 OGLE LMC-ECL-15496 & 82.79 (0.38) &  9200 (fixed) &  8061 (111) & 58.4 (1.3)& 30.4 (0.5)& 11.2 (2.0)& 0.195 (4) & 0.161 (5) & 3563.1201 (67) & 2.1826978 (125)& 0.186 (51)& 231.9 (9.8) &162.0 (55.4) \\ 
 OGLE LMC-ECL-15664 & 82.47 (1.42) & 26000 (fixed) & 16725 (435) & 20.9 (2.4)&  7.3 (1.7)& 71.8 (3.6)& 0.218 (7) & 0.184 (5) & 5600.4875 (65) & 1.8189578 (60) & 0.161 (10)& 330.9 (2.0) & 41.7 (2.3)  \\ 
 OGLE LMC-ECL-15764 & 87.89 (0.59) & 10400 (fixed) &  9442 (209) & 62.1 (1.8)& 34.1 (2.0)& 3.8 (1.7) & 0.185 (5) & 0.144 (3) & 3569.6766 (155)& 2.6937937 (201)& 0.283 (68)&   1.8 (4.9) &152.2 (13.7) \\ 
 OGLE LMC-ECL-15779 & 82.44 (0.41) &  8550 (fixed) &  8109 (252) & 43.8 (0.5)& 27.5 (0.4)& 28.7 (0.6)& 0.205 (3) & 0.168 (3) & 2245.4017 (108)& 2.5036353 (69) & 0.126 (29)& 124.6 (3.8) &132.3 (37.8) \\ 
 OGLE LMC-ECL-16178 & 72.58 (0.35) & 25500 (fixed) & 25320 (263) & 49.7 (1.2)& 46.8 (0.9)& 3.5 (2.2) & 0.260 (4) & 0.255 (2) & 4001.2445 (8)  & 1.6616695 (9)  & 0.023 (3) & 271.8 (0.9) & 14.8 (1.0)  \\ 
 OGLE LMC-ECL-16310 & 81.67 (0.18) & 29000 (fixed) & 27393 (120) & 53.9 (0.6)& 46.1 (0.5)& 0         & 0.234 (2) & 0.228 (2) & 4002.7397 (22) & 2.6313468 (25) & 0.029 (8) & 318.6 (3.1) & 47.7 (2.6)  \\ 
 OGLE LMC-ECL-16320 & 89.73 (0.33) & 23000 (fixed) & 21847 (237) & 56.3 (0.9)& 42.6 (0.7)& 1.1 (1.5) & 0.243 (7) & 0.221 (7) & 4001.5891 (9)  & 2.1091399 (14) & 0.011 (4) &  96.5 (2.5) & 31.2 (4.9)  \\ 
 OGLE LMC-ECL-16414 & 77.48 (0.30) & 17700 (fixed) & 19193 (326) & 29.4 (0.7)& 39.1 (1.4)& 31.5 (2.4)& 0.187 (3) & 0.208 (6) & 4001.2263 (61) & 2.2425933 (115)& 0.125 (19)&  96.6 (4.2) & 51.4 (10.5) \\ 
 OGLE LMC-ECL-16442 & 89.82 (0.77) & 15700 (fixed) & 14992 (394) & 59.0 (0.8)& 41.0 (0.5)& 0         & 0.205 (6) & 0.180 (4) & 4001.8600 (43) & 1.5621158 (48) & 0.178 (28)& 134.6 (3.4) & 70.4 (7.2)  \\ 
 OGLE LMC-ECL-16625 & 85.17 (0.21) & 24500 (fixed) & 23898 (145) & 56.1 (1.6)& 41.6 (1.3)& 2.2 (1.0) & 0.253 (2) & 0.223 (2) & 4001.5240 (11) & 2.6142420 (31) & 0.016 (8) & 229.5 (12.3)& 30.8 (11.0) \\ 
 OGLE LMC-ECL-16732 & 70.56 (0.15) & 24500 (fixed) & 22786 (212) & 42.5 (0.8)& 37.6 (0.7)& 19.9 (2.1)& 0.241 (1) & 0.241 (1) & 4000.7492 (10) & 2.2506837 (21) & 0.025 (9) &  65.4 (7.6) & 38.1 (5.1)  \\ 
 \hline
 \end{tabular}}
\end{table*}

\begin{table*}
 \caption{Parameters of the light curve fits and the apsidal motion, continuation.}
 \label{LCOCparam2}
 \scriptsize
 \centering \scalebox{0.78}{
 \begin{tabular}{l c r c c c c c c c r c c c c c c c}
 \hline\hline
  System            &    $i$       &  \mlc{$T_1$}  & \mlc{$T_2$} &   $L_1$   &   $L_2$   &   $L_3$   &  $R_1/a$  &  $R_2/a$  & $HJD_0     $   &  \mlc{$P$ [d]} &   $e$      & $\omega$ [deg] & $U$ [yr] \\
                    &    [deg]     &  \mlc{[K]}    & \mlc{ [K]}  &   [\%]    &    [\%]   &   [\% ]   &           &           & $[2450000+]$   &                &            &                &          \\
 \hline
 OGLE LMC-ECL-16925 & 86.89 (0.34) & 24500 (fixed) & 21558 (263) & 53.4 (1.6)& 42.7 (1.5)& 3.9 (2.0) & 0.171 (4) & 0.170 (3) & 4001.3216 (61) & 2.3068163 (101)& 0.185 (46) &  63.7 (8.7) & 61.7 (15.8) \\ 
 OGLE LMC-ECL-16928 & 82.78 (0.28) & 10400 (fixed) & 10160 (146) & 48.4 (1.1)& 44.0 (1.0)& 7.6 (3.3) & 0.172 (3) & 0.167 (3) & 4001.3706 (133)& 2.8187346 (206)& 0.245 (66) & 139.6 (11.0)&154.6 (22.0) \\ 
 OGLE LMC-ECL-16983 & 83.96 (0.29) & 14000 (fixed) & 11517 (173) & 59.3 (2.0)& 40.7 (1.4)& 0         & 0.206 (3) & 0.199 (2) & 4001.5687 (27) & 1.6403939 (30) & 0.101 (11) & 121.4 (2.1) & 56.5 (7.2)  \\ 
 OGLE LMC-ECL-17030 & 85.61 (0.52) & 14500 (fixed) & 13369 (317) & 57.1 (1.6)& 38.5 (2.2)& 4.4 (1.7) & 0.194 (5) & 0.170 (7) & 4001.5758 (140)& 2.5185395 (219)& 0.212 (80) & 208.2 (14.6)&114.9 (20.9) \\ 
 OGLE LMC-ECL-17042 & 80.98 (0.27) & 20600 (fixed) & 22193 (224) & 46.6 (0.6)& 53.4 (0.7)& 0         & 0.180 (2) & 0.182 (3) & 4002.4456 (83) & 2.6105475 (135)& 0.200 (34) & 127.8 (4.5) & 80.3 (9.8)  \\ 
 OGLE LMC-ECL-17183 & 83.41 (0.43) & 10400 (fixed) & 11587 (163) & 36.6 (1.0)& 53.0 (1.2)& 11.3 (1.2)& 0.182 (7) & 0.201 (8) & 4001.7230 (57) & 2.2069498 (85) & 0.158 (16) & 133.7 (8.0) &112.6 (7.9)  \\ 
 OGLE LMC-ECL-17198 & 82.48 (0.50) & 24500 (fixed) & 23583 (208) & 29.1 (1.7)& 28.1 (1.3)& 42.8 (2.4)& 0.226 (4) & 0.222 (3) & 5000.1246 (69) & 3.2908144 (224)& 0.079 (13) & 155.3 (5.7) & 74.7 (16.1) \\ 
 OGLE LMC-ECL-17223 & 76.32 (0.38) & 17500 (fixed) & 16835 (531) & 54.7 (1.9)& 37.5 (1.6)& 7.8 (3.0) & 0.170 (5) & 0.144 (5) & 5000.6431 (138)& 2.6021867 (379)& 0.208 (75) & 134.0 (14.3)&104.9 (36.7) \\ 
 OGLE LMC-ECL-17226 & 86.04 (0.45) &  9500 (fixed) &  8802 (193) & 59.3 (1.4)& 40.7 (1.2)& 0         & 0.253 (4) & 0.237 (8) & 6000.6648 (9)  & 0.9879314 (10) & 0.037 (4)  & 319.9 (3.6) & 11.9 (1.0)  \\ 
 OGLE LMC-ECL-17236 & 77.27 (0.43) & 10400 (fixed) &  9976 (309) & 45.9 (1.1)& 30.1 (1.7)& 24.0 (3.5)& 0.226 (3) & 0.183 (3) & 4002.6218 (42) & 1.5755586 (104)& 0.118 (19) &  52.2 (4.1) & 29.6 (3.4)  \\ 
 OGLE LMC-ECL-17267 & 70.88 (0.32) & 25000 (fixed) & 19240 (286) & 43.8 (0.9)& 27.3 (1.4)& 28.9 (1.6)& 0.238 (4) & 0.236 (2) & 5001.2120 (51) & 1.8805590 (52) & 0.032 (5)  &  25.1 (2.7) & 28.1 (1.6)  \\ 
 OGLE LMC-ECL-17283 & 83.26 (0.23) & 15700 (fixed) & 12316 (152) & 54.8 (1.6)& 37.4 (0.8)& 7.8 (2.3) & 0.186 (3) & 0.181 (3) & 4003.4638 (93) & 2.0727585 (160)& 0.227 (43) & 173.4 (5.0) & 59.6 (4.8)  \\ 
 OGLE LMC-ECL-17299 & 83.11 (0.26) & 10400 (fixed) &  9947 (145) & 58.4 (2.3)& 41.6 (1.8)& 0         & 0.194 (5) & 0.174 (4) & 4001.6258 (64) & 1.7431951 (9)  & 0.203 (14) &  72.7 (2.4) & 74.5 (11.2) \\ 
 OGLE LMC-ECL-17406 & 85.17 (0.91) & 21000 (fixed) & 12184 (270) & 38.9 (1.7)&  6.1 (0.5)& 55.0 (3.2)& 0.237 (3) & 0.139 (2) & 5002.9681 (11) & 1.9407060 (34) & 0.120 (72) &  93.8 (13.2)&110.6 (24.1) \\ 
 OGLE LMC-ECL-17498 & 86.01 (0.29) & 16700 (fixed) & 12948 (201) & 57.2 (1.2)& 39.5 (1.0)& 3.3 (1.0) & 0.180 (3) & 0.174 (2) & 4000.1153 (90) & 2.5878412 (162)& 0.251 (12) & 256.0 (6.6) &132.7 (10.7) \\ 
 OGLE LMC-ECL-17543 & 82.12 (0.17) & 12500 (fixed) & 11669 (102) & 57.1 (0.9)& 40.7 (2.0)& 2.2 (1.4) & 0.187 (4) & 0.168 (2) & 4000.0572 (52) & 2.6029786 (90) & 0.106 (20) & 138.8 (3.8) &131.3 (32.0) \\ 
 OGLE LMC-ECL-17579 & 77.23 (0.71) & 24000 (fixed) & 20518 (263) & 32.5 (0.8)& 23.4 (2.7)& 44.1 (4.8)& 0.236 (3) & 0.233 (3) & 4001.1745 (18) & 1.5486575 (44) & 0.041 (17) & 267.4 (13.9)& 72.1 (26.9) \\ 
 OGLE LMC-ECL-17711 & 73.22 (0.10) & 20600 (fixed) & 20232 (179) & 53.1 (0.4)& 46.7 (0.4)& 0.2 (0.2) & 0.227 (2) & 0.217 (2) & 4000.9684 (15) & 1.9613503 (21) & 0.037 (7)  &  71.8 (1.5) & 31.7 (2.3)  \\ 
 OGLE LMC-ECL-17718 & 87.23 (0.38) & 10700 (fixed) &  9978  (99) & 53.0 (0.6)& 47.0 (0.7)& 0         & 0.203 (3) & 0.202 (3) & 5501.6279 (19) & 2.0641028 (19) & 0.039 (14) & 142.5 (9.7) & 66.7 (11.4) \\ 
 OGLE LMC-ECL-17774 & 79.35 (0.23) & 25500 (fixed) & 23836 (333) & 32.2 (1.7)& 29.8 (1.3)& 38.0 (3.5)& 0.188 (4) & 0.174 (3) & 5798.4747 (47) & 3.9294300 (173)& 0.070 (13) & 319.7 (3.4) &  7.2 (0.2)  \\ 
 OGLE LMC-ECL-17777 & 86.17 (0.31) & 20600 (fixed) & 20376 (176) & 52.4 (0.9)& 47.0 (1.1)& 0.6 (0.6) & 0.199 (2) & 0.195 (2) & 4002.2267 (91) & 2.8638071 (160)& 0.125 (11) &  35.8 (6.1) & 87.3 (7.8)  \\ 
 OGLE LMC-ECL-17800 & 74.26 (0.14) & 17700 (fixed) & 11632 (188) & 62.8 (1.3)& 28.7 (0.9)& 8.5 (1.3) & 0.237 (4) & 0.222 (3) & 4001.9965 (12) & 1.4354526 (12) & 0.035 (12) & 206.8 (5.0) & 20.5 (1.7)  \\ 
 OGLE LMC-ECL-17809 & 87.98 (0.32) & 10400 (fixed) & 10551  (97) & 47.9 (1.0)& 52.1 (0.9)& 0         & 0.210 (2) & 0.213 (3) & 4002.3658 (49) & 2.2079038 (85) & 0.111 (30) &  65.0 (3.8) & 70.2 (4.0)  \\ 
 OGLE LMC-ECL-17923 & 89.53 (0.49) & 15700 (fixed) & 15628 (129) & 52.9 (0.5)& 47.1 (0.4)& 0         & 0.208 (2) & 0.196 (2) & 4002.8307 (77) & 2.5538851 (145)& 0.150 (27) & 145.7 (5.5) & 82.6 (14.1) \\ 
 OGLE LMC-ECL-18077 & 82.54 (0.60) & 10700 (fixed) & 12604 (300) & 39.4 (0.6)& 52.9 (1.8)& 7.7 (1.6) & 0.159 (5) & 0.165 (6) & 5002.2916 (31) & 1.7835501 (88) & 0.221 (98) &  88.5 (18.7)&112.9 (35.6) \\ 
 OGLE LMC-ECL-18152 & 78.24 (0.29) & 26000 (fixed) & 22707 (328) & 64.1 (0.9)& 32.3 (1.2)& 3.6 (2.0) & 0.208 (4) & 0.168 (4) & 5001.9415 (47) & 2.5629074 (168)& 0.098 (22) &  98.3 (7.7) & 45.5 (8.2)  \\ 
 OGLE LMC-ECL-18174 & 79.75 (0.15) & 10700 (fixed) & 11510 (125) & 44.3 (0.4)& 55.7 (0.4)& 0         & 0.207 (2) & 0.219 (2) & 4003.0395 (39) & 2.5366697 (63) & 0.129 (11) & 123.2 (2.9) & 81.7 (7.0)  \\ 
 OGLE LMC-ECL-18316 & 89.70 (0.47) & 17000 (fixed) & 16256 (190) & 62.5 (1.3)& 35.2 (1.1)& 2.3 (1.4) & 0.226 (3) & 0.171 (3) & 5001.7243 (80) & 2.1080445 (202)& 0.166 (30) & 103.1 (5.1) & 24.1 (3.8)  \\ 
 OGLE LMC-ECL-18355 & 79.65 (0.58) & 21000 (fixed) & 12642 (509) & 44.0 (1.6)&  7.6 (0.8)& 48.4 (3.2)& 0.233 (4) & 0.139 (9) & 5002.0211 (41) & 2.1675189 (126)& 0.141 (19) & 250.0 (4.2) & 45.1 (9.9)  \\ 
 OGLE LMC-ECL-18371 & 81.36 (0.25) & 14000 (fixed) & 14885 (197) & 36.9 (0.6)& 63.1 (0.6)& 0         & 0.186 (3) & 0.234 (4) & 5001.9229 (24) & 1.7069521 (25) & 0.056 (11) &  91.0 (3.6) & 48.6 (18.7) \\ 
 OGLE LMC-ECL-18408 & 81.27 (0.19) & 10600 (fixed) & 11018 (162) & 47.8 (0.3)& 52.2 (0.3)& 0         & 0.157 (2) & 0.159 (2) & 4002.5205 (85) & 2.3382018 (253)& 0.151 (52) & 137.1 (14.9)&142.2 (72.6) \\ 
 OGLE LMC-ECL-18500 & 82.08 (0.25) & 14500 (fixed) & 14602 (208) & 47.9 (0.3)& 52.1 (0.4)& 0         & 0.199 (3) & 0.207 (4) & 5002.2085 (63) & 2.3546062 (110)& 0.166 (46) &  79.6 (21.0)&151.7 (77.8) \\ 
 OGLE LMC-ECL-18501 & 85.26 (0.33) & 13000 (fixed) & 15439 (457) & 40.0 (2.2)& 53.5 (3.2)& 6.5 (2.7) & 0.148 (4) & 0.151 (4) & 5002.6321 (87) & 2.3935782 (121)& 0.290 (109)&  97.0 (11.5)&149.2 (82.0) \\ 
 OGLE LMC-ECL-18579 & 80.73 (0.44) & 14500 (fixed) & 12708 (346) & 63.1 (2.4)& 32.3 (4.0)& 4.6 (1.1) & 0.163 (3) & 0.130 (7) & 4002.1199 (160)& 3.1174213 (199)& 0.301 (43) & 145.8 (7.3) &105.6 (16.1) \\ 
 OGLE LMC-ECL-19095 & 81.90 (0.17) & 24500 (fixed) & 30371 (232) & 39.7 (0.6)& 58.5 (1.3)& 1.8 (0.8) & 0.169 (2) & 0.170 (3) & 4002.1948 (151)& 2.9751066 (81) & 0.249 (94) & 126.5 (13.4)&103.7 (22.3) \\ 
 OGLE LMC-ECL-19294 & 84.00 (0.30) & 18500 (fixed) &  8067 (169) & 70.7 (1.2)& 29.3 (1.0)& 0         & 0.220 (3) & 0.172 (3) & 4003.8965 (82) & 2.7363868 (138)& 0.148 (27) & 245.6 (6.1) &100.2 (7.5)  \\ 
 OGLE LMC-ECL-19577 & 73.06 (0.22) & 20600 (fixed) & 22170 (214) & 35.7 (0.6)& 64.3 (0.7)& 0         & 0.220 (2) & 0.277 (3) & 3006.7611 (13) & 1.6514537 (16) & 0.051 (4)  & 204.5 (7.0) & 12.8 (1.0)  \\ 
 OGLE LMC-ECL-19583 & 85.91 (0.42) & 24500 (fixed) & 20917 (351) & 50.2 (2.0)& 38.4 (1.6)& 11.4 (2.5)& 0.175 (3) & 0.173 (3) & 4002.6050 (118)& 2.1894745 (130)& 0.212 (18) & 218.2 (5.2) & 99.0 (8.2)  \\ 
 OGLE LMC-ECL-19624 & 81.27 (0.33) & 26000 (fixed) & 24418 (217) & 59.3 (1.7)& 40.7 (1.9)& 0         & 0.193 (3) & 0.164 (2) & 3002.2117 (155)& 2.7015912 (227)& 0.279 (25) &  40.9 (3.0) & 50.9 (4.2)  \\ 
 OGLE LMC-ECL-19675 & 89.54 (0.59) & 12500 (fixed) & 12103 (260) & 50.7 (1.2)& 48.1 (0.8)& 1.2 (1.1) & 0.178 (4) & 0.178 (4) & 4003.1777 (53) & 2.7038366 (203)& 0.149 (61) & 108.1 (4.7) &172.5 (19.4) \\ 
 OGLE LMC-ECL-19727 & 84.50 (0.16) & 12500 (fixed) & 18828 (370) & 26.2 (0.7)& 73.7 (0.9)& 0         & 0.129 (2) & 0.160 (3) & 3007.0991 (71) & 2.6668363 (215)& 0.350 (110)&  77.9 (8.3) &158.8 (45.1) \\ 
 OGLE LMC-ECL-19879 & 86.96 (0.48) & 29000 (fixed) & 28978 (305) & 49.3 (0.9)& 37.4 (1.5)& 13.2 (3.0)& 0.187 (2) & 0.166 (6) & 2282.2391 (127)& 3.8818771 (278)& 0.219 (19) &  96.3 (6.6) &103.4 (16.7) \\ 
 OGLE LMC-ECL-20053 & 75.59 (0.63) & 20600 (fixed) & 20207 (193) & 47.9 (1.1)& 41.2 (0.8)& 10.9 (2.4)& 0.234 (7) & 0.219 (5) & 4002.0741 (125)& 1.9068881 (22) & 0.038 (11) &  80.2 (2.9) & 25.0 (1.9)  \\ 
 OGLE LMC-ECL-20066 & 83.01 (0.26) & 17000 (fixed) & 12812 (166) & 65.7 (0.6)& 34.3 (0.4)& 0         & 0.188 (5) & 0.161 (5) & 5004.1508 (81) & 2.4753628 (198)& 0.209 (38) & 269.7 (4.3) & 50.9 (18.8) \\ 
 OGLE LMC-ECL-20285 & 84.78 (0.17) & 10000 (fixed) &  9496 (111) & 51.3 (1.8)& 48.7 (1.6)& 0         & 0.249 (2) & 0.246 (2) & 4001.6069 (15) & 1.7624139 (33) & 0.031 (8)  &  91.9 (1.7) & 17.5 (1.5)  \\ 
 OGLE LMC-ECL-20299 & 79.80 (0.25) & 31500 (fixed) & 32385 (136) & 42.7 (0.4)& 46.8 (0.9)& 10.5 (1.5)& 0.210 (3) & 0.216 (2) & 4002.8032 (23) & 2.8414568 (43) & 0.053 (7)  &  48.9 (2.2) & 41.4 (5.0)  \\ 
 OGLE LMC-ECL-20384 & 83.55 (0.48) & 26000 (fixed) & 15442 (242) & 84.5 (1.1)& 15.5 (1.0)& 0         & 0.222 (2) & 0.147 (1) & 4001.1584 (72) & 2.4038316 (129)& 0.165 (10) & 196.7 (4.1) & 54.6 (4.8)  \\ 
 OGLE LMC-ECL-20550 & 88.83 (0.69) & 13000 (fixed) & 12758 (138) & 55.1 (1.8)& 42.5 (3.1)& 2.4 (2.4) & 0.249 (6) & 0.224 (4) & 4000.8871 (10) & 1.4474212 (5)  & 0.040 (5)  & 347.8 (1.0) & 13.4 (1.2)  \\ 
 OGLE LMC-ECL-20589 & 83.75 (0.21) & 26000 (fixed) & 24274 (226) & 54.2 (0.7)& 45.8 (0.5)& 0         & 0.195 (3) & 0.190 (2) & 2926.6720 (133)& 3.8535274 (326)& 0.195 (33) & 207.1 (9.9) &108.8 (4.5)  \\ 
 OGLE LMC-ECL-20648 & 82.60 (0.15) & 17500 (fixed) & 18913 (230) & 42.3 (0.6)& 57.6 (0.8)& 0         & 0.169 (3) & 0.186 (3) & 4002.2779 (131)& 2.5286693 (35) & 0.241 (12) & 196.5 (7.6) & 67.5 (8.2)  \\ 
 OGLE LMC-ECL-20735 & 83.37 (0.40) &  9500 (fixed) &  9167 (175) & 55.6 (0.4)& 44.4 (0.3)& 0         & 0.223 (5) & 0.208 (3) & 4003.5952 (18) & 1.7966394 (23) & 0.049 (16) &  57.0 (2.4) & 58.6 (4.0)  \\ 
 OGLE LMC-ECL-20783 & 87.99 (0.17) & 24000 (fixed) & 24253 (120) & 46.7 (0.6)& 52.3 (0.5)& 1.0 (1.0) & 0.236 (2) & 0.248 (3) & 9184.2211 (70) & 3.2566811 (80) & 0.082 (13) & 319.5 (3.5) & 60.2 (6.1)  \\ 
 OGLE LMC-ECL-20860 & 83.87 (0.85) & 17000 (fixed) & 13737 (191) & 33.7 (1.8)& 20.9 (2.4)& 45.4 (5.6)& 0.229 (7) & 0.210 (5) & 4002.8836 (27) & 1.6441828 (33) & 0.049 (11) & 253.8 (2.2) & 22.1 (2.3)  \\ 
 OGLE LMC-ECL-21259 & 80.40 (0.76) & 25500 (fixed) & 13838 (334) & 54.0 (3.2)&  9.1 (2.7)& 36.9 (6.1)& 0.235 (4) & 0.155 (6) & 4001.1459 (78) & 2.5508208 (251)& 0.133 (30) & 229.3 (6.0) & 73.6 (23.4) \\ 
 OGLE LMC-ECL-21278 & 86.90 (0.99) & 26000 (fixed) & 19842 (342) & 62.6 (2.5)& 20.7 (0.8)& 16.7 (3.7)& 0.255 (6) & 0.178 (3) & 4002.6536 (13) & 1.6617552 (27) & 0.039 (5)  &  77.1 (2.6) & 31.1 (7.2)  \\ 
 OGLE LMC-ECL-21479 & 79.09 (0.60) & 10400 (fixed) &  8006 (207) & 69.9 (2.4)& 30.1 (2.2)& 0         & 0.211 (6) & 0.182 (9) & 5550.6183 (99) & 2.0602435 (244)& 0.177 (41) & 160.3 (7.1) & 37.7 (8.5)  \\ 
 OGLE LMC-ECL-21647 & 74.07 (0.21) & 17000 (fixed) & 17390 (156) & 48.9 (0.4)& 51.1 (0.4)& 0         & 0.221 (3) & 0.221 (3) & 4001.3485 (36) & 2.6411620 (79) & 0.084 (20) & 127.6 (4.5) & 77.3 (8.7)  \\ 
 OGLE LMC-ECL-21695 & 81.88 (0.18) & 17000 (fixed) & 14154 (121) & 57.8 (1.0)& 42.2 (0.9)& 0         & 0.177 (2) & 0.171 (3) & 4001.6820 (122)& 2.7611224 (396)& 0.200 (45) & 227.1 (8.4) &154.4 (15.2) \\ 
 OGLE LMC-ECL-21961 & 89.74 (0.37) & 16000 (fixed) & 14973 (218) & 56.4 (0.3)& 43.6 (0.3)& 0         & 0.194 (2) & 0.179 (2) & 4001.0849 (42) & 2.0607986 (98) & 0.096 (6)  & 296.4 (3.2) & 74.3 (5.7)  \\ 
 OGLE LMC-ECL-21968 & 75.48 (0.71) & 24500 (fixed) & 29213 (315) & 27.9 (0.9)& 60.0 (1.3)& 12.1 (1.8)& 0.175 (3) & 0.218 (5) & 4001.6143 (31) & 2.8824655 (130)& 0.150 (52) &  79.6 (6.2) & 69.6 (30.9) \\ 
 OGLE LMC-ECL-22069 & 86.52 (0.27) & 31500 (fixed) & 29685 (276) & 52.9 (1.2)& 44.7 (1.4)& 2.4 (1.3) & 0.174 (4) & 0.170 (2) & 4000.7189 (49) & 2.2719812 (159)& 0.210 (31) & 135.6 (3.3) & 78.1 (15.8) \\ 
 OGLE LMC-ECL-22211 & 80.34 (0.40) & 16000 (fixed) & 13995 (295) & 58.1 (0.8)& 41.9 (0.8)& 0         & 0.244 (2) & 0.228 (3) & 4000.8583 (19) & 1.3846511 (18) & 0.043 (4)  & 275.8 (1.8) & 17.3 (2.0)  \\ 
 OGLE LMC-ECL-22232 & 80.29 (0.48) & 20600 (fixed) & 19708 (149) & 54.8 (1.4)& 30.2 (0.9)& 15.0 (3.2)& 0.228 (4) & 0.176 (3) & 4000.7192 (36) & 2.0099807 (57) & 0.090 (25) & 323.2 (4.5) & 36.2 (6.6)  \\ 
 OGLE LMC-ECL-22422 & 85.19 (0.36) & 13500 (fixed) & 12558 (220) & 52.4 (1.1)& 47.6 (1.0)& 0         & 0.216 (3) & 0.218 (3) & 4001.2210 (20) & 1.5616800 (46) & 0.069 (18) &  45.5 (6.1) & 43.0 (11.3) \\ 
 OGLE LMC-ECL-22455 & 79.54 (0.49) & 20600 (fixed) & 23856 (478) & 22.1 (0.7)& 34.8 (0.6)& 43.1 (4.5)& 0.172 (3) & 0.192 (7) & 4000.1400 (22) & 1.9563996 (37) & 0.159 (41) & 101.2 (3.2) & 67.0 (13.1) \\ 
 OGLE LMC-ECL-22494 & 83.08 (0.42) & 14000 (fixed) & 14492 (298) & 41.8 (0.6)& 57.8 (0.5)& 0.4 (0.4) & 0.199 (4) & 0.229 (6) & 4001.3256 (13) & 1.3164150 (14) & 0.064 (8)  & 256.6 (1.9) & 45.9 (8.2)  \\ 
 OGLE LMC-ECL-22613 & 75.84 (0.53) & 30000 (fixed) & 26863 (303) & 46.8 (1.0)& 25.1 (2.1)& 28.1 (2.8)& 0.281 (4) & 0.232 (6) & 4000.2078 (21) & 1.7859888 (32) & 0.010 (3)  &  70.9 (2.8) &  6.6 (0.3)  \\ 
 OGLE LMC-ECL-22695 & 78.20 (0.18) & 22000 (fixed) & 22550 (127) & 43.2 (0.5)& 56.8 (0.5)& 0         & 0.204 (2) & 0.229 (2) & 4002.2386 (23) & 2.3403132 (41) & 0.050 (9)  & 105.5 (3.4) & 37.8 (7.5)  \\ 
 OGLE LMC-ECL-23298 & 76.82 (0.29) & 26000 (fixed) & 22947 (311) & 41.0 (1.2)& 31.4 (0.8)& 27.6 (3.1)& 0.278 (5) & 0.272 (5) & 4000.8389 (12) & 1.9329663 (13) & 0.012 (4)  &  63.6 (1.3) & 12.2 (0.9)  \\ 
 OGLE LMC-ECL-23323 & 80.96 (0.35) & 15700 (fixed) & 11814 (143) & 56.1 (0.9)& 36.2 (0.6)& 7.7 (1.9) & 0.237 (3) & 0.236 (3) & 4001.0205 (13) & 1.3528091 (19) & 0.034 (10) & 275.9 (3.2) & 20.1 (6.7)  \\ 
 OGLE LMC-ECL-23920 & 81.51 (0.48) & 17800 (fixed) & 13018 (286) & 46.5 (1.7)& 24.6 (1.0)& 28.9 (3.2)& 0.226 (6) & 0.210 (5) & 4001.1443 (29) & 1.8619509 (71) & 0.083 (17) &  92.5 (4.0) & 46.6 (14.9) \\ 
 OGLE LMC-ECL-24112 & 76.95 (0.29) & 15700 (fixed) & 19418 (400) & 24.3 (1.1)& 49.1 (1.5)& 26.6 (2.0)& 0.176 (3) & 0.214 (5) & 5002.0329 (71) & 3.5598499 (180)& 0.150 (42) & 120.4 (5.5) &126.4 (32.6) \\ 
 OGLE LMC-ECL-24236 & 79.67 (0.37) & 15000 (fixed) & 10398 (328) & 57.4 (2.3)& 14.0 (1.2)& 28.6 (1.7)& 0.230 (3) & 0.164 (3) & 5002.4655 (34) & 1.6278800 (85) & 0.118 (38) & 250.1 (3.7) & 55.2 (14.8) \\ 
 OGLE LMC-ECL-24534 & 89.75 (0.19) & 29000 (fixed) & 26600 (129) & 60.8 (0.3)& 39.2 (0.3)& 0         & 0.210 (3) & 0.183 (2) & 5000.8369 (20) & 2.4958955 (64) & 0.094 (31) &  82.5 (1.9) & 60.6 (19.0) \\ 
 OGLE LMC-ECL-24817 & 84.16 (0.46) & 11000 (fixed) & 11671 (206) & 43.7 (2.2)& 51.2 (1.6)& 5.1 (1.9) & 0.184 (5) & 0.191 (3) & 5002.1038 (77) & 2.1453406 (172)& 0.165 (48) & 129.7 (6.4) &107.2 (42.1) \\ 
 OGLE LMC-ECL-25047 & 86.08 (0.49) & 20600 (fixed) & 20516 (317) & 55.3 (3.0)& 38.9 (2.1)& 5.8 (3.2) & 0.268 (4) & 0.208 (7) & 5000.6412 (10) & 1.0137603 (11) & 0.049 (12) &   5.8 (2.1) & 10.8 (1.5)  \\ 
 OGLE LMC-ECL-25227 & 85.50 (0.30) & 17000 (fixed) & 14237 (299) & 68.1 (0.5)& 31.9 (0.5)& 0         & 0.206 (3) & 0.160 (3) & 5002.3763 (59) & 1.9394433 (92) & 0.168 (29) & 257.0 (4.0) & 78.9 (7.2)  \\ 
 OGLE LMC-ECL-25743 & 83.61 (0.17) & 26000 (fixed) & 25326 (105) & 54.5 (0.6)& 45.5 (0.4)& 0         & 0.252 (3) & 0.241 (3) & 5000.9462 (62) & 3.8107354 (320)& 0.111 (38) & 114.1 (4.8) & 83.6 (12.9) \\ 
 OGLE LMC-ECL-25885 & 85.69 (0.22) & 10000 (fixed) & 10186 (156) & 49.0 (0.4)& 51.0 (0.4)& 0         & 0.199 (3) & 0.201 (4) & 5001.3961 (45) & 3.2780824 (129)& 0.193 (77) & 103.9 (11.2)&156.9 (29.8) \\ 
 OGLE LMC-ECL-25980 & 81.58 (0.41) & 17000 (fixed) & 15824 (120) & 55.6 (1.1)& 42.2 (0.9)& 2.2 (1.7) & 0.262 (3) & 0.239 (5) & 5001.8010 (10) & 1.7887515 (28) & 0.028 (3)  &  83.4 (3.9) & 30.8 (4.5)  \\ 
 \hline
 \end{tabular}}
\end{table*}

\begin{table*}
\caption{Third-body orbit parameters of the individual systems.} \label{LITEparam}
\small
\begin{tabular}{cccccc|ccc}
\hline\hline\noalign{\smallskip}
  System           &   $A$       & $\omega_3$   & $P_3$      & $T_0$ [HJD] &    $e_3$    & $f(m_3)$   & $P^2/P_3$  & $P_3^2/P$  \\   
                   &  [days]     &   [deg]      & [yr]       & (2400000+)  &             & $[M_\odot]$& [days]     & [yr]       \\   
 \noalign{\smallskip}\hline\noalign{\smallskip}
 OGLE LMC-ECL-00483& 0.0101 (23) & 186.9 (15.1) & 16.3 (1.7) & 50041 (5400)& 0.464 (96)  & 0.029 (10) & 0.0006377 & 50364      \\   
 OGLE LMC-ECL-00737& 0.0055 (17) & 150.0 (18.0) &  8.7 (1.1) & 51660 (1809)& 0.449 (21)  & 0.014 (3)  & 0.0012394 & 13951      \\   
 OGLE LMC-ECL-02912& 0.0054 (12) &  76.1 (11.4) &  7.9 (1.3) & 59187 (239) & 0.484 (57)  & 0.014 (3)  & 0.0015495 & 10722      \\   
 OGLE LMC-ECL-07578& 0.0051 (7)  &  61.6 (12.3) & 43.4 (7.4) & 51528 (481) & 0.450 (103) & 0.390 (89) & 0.0010190 & 171095     \\   
 OGLE LMC-ECL-10302& 0.0058 (4)  & 148.2 (22.1) &  6.5 (0.3) & 66252 (164) & 0.079 (26)  & 0.025 (2)  & 0.0010042 & 9917       \\   
 OGLE LMC-ECL-11351& 0.0121 (9)  & 244.4 (14.9) &  8.7 (0.5) & 67153 (872) & 0.080 (30)  & 0.122 (23) & 0.0018351 & 11390      \\   
 OGLE LMC-ECL-11854& 0.0061 (14) &  18.1 (23.7) & 16.3 (6.6) & 57787 (1289)& 0.669 (134) & 0.010 (24) & 0.0009691 & 40305      \\   
 OGLE LMC-ECL-12043& 0.0114 (18) & 130.4 (16.0) &  6.1 (0.9) & 65487 (578) & 0.121 (112) & 0.207 (17) & 0.0008039 & 10268      \\   
 OGLE LMC-ECL-12504& 0.0042 (17) & 129.1 (24.8) &  7.7 (4.0) & 55933 (1302)& 0.114 (88)  & 0.006 (4)  & 0.0020258 & 9077       \\   
 OGLE LMC-ECL-12792& 0.0138 (24) &  26.5 (20.0) & 18.7 (1.3) & 53958 (375) & 0.087 (43)  & 0.039 (3)  & 0.0004409 & 73625      \\   
 OGLE LMC-ECL-15161& 0.0035 (16) &  71.4 (18.6) &  7.2 (0.8) & 65075 (1235)& 0.061 (89)  & 0.004 (2)  & 0.0050121 & 5193       \\   
 OGLE LMC-ECL-15288& 0.0065 (10) & 256.5 (31.8) & 14.2 (3.0) & 67590 (772) & 0.051 (11)  & 0.007 (2)  & 0.0013768 & 27628      \\   
 OGLE LMC-ECL-15473& 0.0055 (9)  & 179.4 (10.5) &  3.6 (0.2) & 65832 (198) & 0.077 (23)  & 0.066 (11) & 0.0019212 & 2977       \\   
 OGLE LMC-ECL-15664& 0.0244 (11) & 158.1 (8.2)  &  4.9 (0.3) & 58687 (240) & 0.431 (20)  & 4.123 (127)& 0.0018504 & 4812       \\   
 OGLE LMC-ECL-15779& 0.0118 (23) & 192.8 (32.8) &  3.4 (0.9) & 58062 (181) & 0.001 (13)  & 0.733 (86) & 0.0050449 & 1688       \\   
 OGLE LMC-ECL-16320& 0.0118 (12) & 170.1 (7.4)  & 20.1 (1.3) & 62128 (523) & 0.372 (82)  & 0.027 (6)  & 0.0006066 & 69760      \\   
 OGLE LMC-ECL-16414& 0.0199 (16) &  28.8 (18.2) & 12.8 (2.5) & 69372 (526) & 0.069 (14)  & 0.252 (72) & 0.0010706 & 26940      \\   
 OGLE LMC-ECL-16928& 0.0108 (24) &  94.3 (11.1) & 16.6 (4.1) & 67522 (820) & 0.094 (32)  & 0.024 (18) & 0.0013116 & 35643      \\   
 OGLE LMC-ECL-17183& 0.0060 (9)  & 166.3 (10.9) & 10.1 (2.7) & 67869 (533) & 0.069 (13)  & 0.011 (5)  & 0.0013146 & 17029      \\   
 OGLE LMC-ECL-17198& 0.0102 (17) & 191.3 (8.4)  &  2.9 (0.3) & 89367 (114) & 0.599 (65)  & 1.218 (79) & 0.0100834 & 959        \\   
 OGLE LMC-ECL-17236& 0.0079 (10) & 305.5 (15.2) &  8.4 (3.7) & 49564 (1270)& 0.317 (115) & 0.037 (12) & 0.0008057 & 16493      \\   
 OGLE LMC-ECL-17267& 0.0517 (110)& 300.3 (20.1) & 42.9 (8.9) & 85049 (1431)& 0.749 (143) & 0.490 (139)& 0.0002257 & 357586     \\   
 OGLE LMC-ECL-17498& 0.0042 (9)  & 209.4 (16.0) & 10.2 (2.2) & 65941 (418) & 0.188 (72)  & 0.004 (1)  & 0.0018049 & 14573      \\   
 OGLE LMC-ECL-17579& 0.0193 (22) &  56.8 (17.4) &  7.8 (0.6) & 65679 (172) & 0.187 (14)  & 0.631 (19) & 0.0008424 & 14321      \\   
 OGLE LMC-ECL-17774& 0.0409 (26) &  10.4 (10.2) & 11.1 (1.1) & 52695 (370) & 0.378 (72)  & 3.605 (82) & 0.0037569 & 11456      \\   
 OGLE LMC-ECL-17800& 0.0055 (11) & 251.4 (32.7) & 14.8 (2.4) & 67039 (696) & 0.203 (83)  & 0.004 (2)  & 0.0003806 & 55890      \\   
 OGLE LMC-ECL-18355& 0.0118 (16) & 359.9 (11.6) &  4.9 (1.0) & 64531 (389) & 0.226 (68)  & 0.392 (90) & 0.0026393 & 4002       \\   
 OGLE LMC-ECL-18579& 0.0062 (23) & 194.6 (22.5) &  9.3 (1.8) & 63356 (665) & 0.211 (110) & 0.016 (9)  & 0.0028625 & 10123      \\   
 OGLE LMC-ECL-19583& 0.0057 (17) & 358.3 (10.1) &  7.9 (1.1) & 64448 (402) & 0.126 (60)  & 0.016 (3)  & 0.0016531 & 10515      \\   
 OGLE LMC-ECL-19879& 0.0096 (8)  & 253.3 (8.9)  &  7.0 (0.5) & 58977 (211) & 0.747 (101) & 0.101 (66) & 0.0056815 & 4629       \\   
 OGLE LMC-ECL-20053& 0.0033 (11) & 236.1 (21.8) &  5.6 (1.0) & 67592 (326) & 0.209 (93)  & 0.006 (3)  & 0.0017783 & 6003       \\   
 OGLE LMC-ECL-20299& 0.0039 (7)  & 112.3 (10.4) &  9.7 (1.6) & 63550 (501) & 0.141 (76)  & 0.003 (2)  & 0.0022804 & 12180      \\   
 OGLE LMC-ECL-20860& 0.0437 (9)  &   1.4 (8.5)  & 19.7 (0.8) & 63450 (234) & 0.107 (13)  & 1.143 (15) & 0.0003763 & 85949      \\   
 OGLE LMC-ECL-21968& 0.0063 (11) & 355.7 (12.8) &  3.7 (0.7) & 54143 (290) & 0.750 (109) & 0.322 (77) & 0.0060234 & 1762       \\   
 OGLE LMC-ECL-22232& 0.0049 (8)  &   3.7 (13.0) &  8.8 (1.6) & 55716 (498) & 0.749 (152) & 0.027 (11) & 0.0012493 & 14245      \\   
 OGLE LMC-ECL-22613& 0.0449 (12) & 189.2 (3.7)  & 23.5 (0.4) & 57134 (141) & 0.521 (21)  & 1.344 (48) & 0.0003709 & 113408     \\   
  \hline
 \noalign{\smallskip}\hline
\end{tabular}
\end{table*}

 \end{appendix}

\end{document}